\documentclass[useAMS,usenatbib]{mn2e}
\usepackage{color}
\usepackage{graphicx}

\newcommand{\apj}{ApJ}                                        
\newcommand{\aap}{A\&A}                                       
\newcommand{\aaps}{Astronomy and Astrophysics Supplement}     
                                      
\newcommand{\apjs}{ApJS}  
\newcommand{\mnras}{MNRAS}                                    

\title[NGC 2355]{
Photometric and spectroscopic study of the intermediate-age open cluster NGC~2355
\thanks{Based on observations 
collected at the Large Binocular Telescope
(LBT) and the Nordic Optical Telescope (NOT). The LBT is an international collaboration among institutions in the
United States, Italy and Germany. LBT Corporation partners are: The University
of Arizona on behalf of the Arizona University system; Istituto Nazionale di
Astrofisica, Italy; LBT Beteiligungsgesellschaft, Germany, representing the
Max-Planck Society, the Astrophysical Institute Potsdam, and Heidelberg
University; The Ohio State University, and The Research Corporation, on behalf
of The University of Notre Dame, University of Minnesota and University of
Virginia. The NOT is operated by the Nordic Optical Telescope Scientific Association at the Observatorio del Roque de los Muchachos, La Palma, Spain, of the Instituto de Astrofisica de Canarias.}}
\author[Donati et al.]{P. Donati$^{1,2}$, A. Bragaglia$^{1}$,  E. Carretta$^{1}$, V. D'Orazi$^{3,4,5}$, M. Tosi$^{1}$, F. Cusano$^{1}$, \newauthor R. Carini$^{6}$
\\
$^{1}$INAF-Osservatorio Astronomico di Bologna, via Ranzani 1, I-40127 Bologna, Italy\\
$^{2}$Dipartimento di Fisica e Astronomia, via Ranzani 1, I-40127 Bologna, Italy\\
$^{3}$INAF-Osservatorio Astronomico di Padova, vicolo Osservatorio 5, I-35122 Padova, Italy\\
$^{4}$Department of Physics and Astronomy, Macquarie University, Sydney, NSW 2109, Australia\\
$^{5}$Monash Centre for Astrophysics, School of Physics and Astronomy, Monash University, Melbourne, VIC 3800, Australia \\
$^{6}$INAF-Osservatorio Astronomico di Roma, via Frascati 33, I-00078, Monte Porzio Catone, Italy}

\begin{document}

\pagerange{\pageref{firstpage}--\pageref{lastpage}} 

\maketitle

\label{firstpage}

\begin{abstract}
In this paper we analyse the evolutionary status and
properties of the old open cluster NGC~2355, located in the
Galactic anticentre direction, as a part of the long term programme BOCCE. NGC~2355 was observed
with LBC@LBT using the Bessel $B$, $V$, and $I_c$ filters. The
cluster parameters have been obtained using the synthetic colour-magnitude
diagram (CMD) method, as done in other papers of this series. 
Additional spectroscopic observations with FIES@NOT of three
giant stars were used to determine the chemical properties of the
cluster. Our analysis shows that NGC~2355 has metallicity slightly less than solar, with [Fe/H]$=-0.06$ dex, age between 0.8 and 1 Gyr, reddening $E(B-V)$ in the range 0.14 and 0.19 mag, and distance modulus $(m-M)_0$ of about 11 mag. We also investigated the abundances of O, Na, Al, $\alpha$, iron-peak, and neutron capture elements, showing that NGC~2355 falls within the abundance distribution of similar clusters (same age and metallicity). The Galactocentric distance of NGC~2355 places it at the border between two regimes of metallicity distribution; this makes it an important cluster for the study of the chemical properties and evolution of the disc. 
\end{abstract}

\section{Introduction}\label{sec:intro}
This paper is part of the BOCCE (Bologna Open Clusters Chemical Evolution)
project, described in detail by \cite{boc_06}. BOCCE aims to precisely and
homogeneously derive the fundamental properties of a large, significant sample
of open clusters (OCs), which are among the best tracers of the Galactic disc
properties (e.g., \citealt{fri_95}).  We have already analysed photometric data
for 35 OCs \citep[see][for updated references]{lbt2}, by comparing observed and
synthetic colour-magnitude diagrams (CMDs, see \citealt{tosi91,boc_06}) and
producing age, distance, reddening, and approximate metallicity on a homogenous
scale.  Metallicity and detailed chemical  abundances based on high-resolution
spectroscopy are instead available for about one third of the sample,  see e.g.
\cite*{carretta07}, and \cite{andreuzzi} for a discussion on the  use of OCs to
ascertain the metallicity distribution in the disc.
\begin{table*}
\centering
\caption{Properties of NGC~2355 in literature sources. If not given in the original paper, $(m-M)_0$ was computed from
$(m-M)_V$ and $E(B-V)$, or from the distance from the Sun.
}
\begin{tabular}{lcccccccl}
\hline
Paper & tel./instr. & FoV/stars & phot/spec & age & $(m-M)_0$ & $E(B-V)$ & [Fe/H] & Notes \\
      &             &  &  &(Gyr) & &  & & \\
\hline
Kaluzny\&Mazur 1991  &0.9m KPNO & $6.6'\times6.6'$ &$UBV$ & ``Praesepe'' & 11.73 & 0.12 & +0.13 & \\
Ann et al. 1999      &1.8m BOAO &$11.8'\times11.8'$ &$UBVI$ & 1 & 11.4 & 0.25 &  --0.32 & Padova isoc.\\
Oliveira et al. 2013 & & & & 0.8 & 11.45 & 0.22 & --0.23 & Using KM91\\ 
                     & & & & 0.9 & 10.88 & 0.32 & --0.32 & Using A99 \\ 
Soubiran et al. 2000 &ELODIE & 24 stars &R=42000 & 1 & 11.06 & 0.16 & --0.07 & 17 members  \\
Jacobson et al. 2011 &Hydra  & 12 stars &R=18000 & & & & --0.08& 6 members\\
\hline
\end{tabular}
\label{t:lit}
\end{table*}

We present here a photometric and spectroscopic analysis of NGC~2355, an
intermediate-age OC located towards  the Galactic anticentre ($l=203.390\deg$;
$b=11.803\deg$, \citealt{dias_02}). The cluster has been targeted  because of
its location in the outer disc ($R_{GC}\ga10$ kpc), close to the point where the
metallicity distribution  seems to flatten, and where only a few OCs have been
studied so far. Furthermore, its stars are bright enough  to be easily observed
with high-resolution spectroscopy.

There are only two previous sources of visual CCD photometry for NGC~2355, both
acquired at relatively small telescopes (KPNO and BOAO). \cite{km91} obtained
$UBV$ data, while \cite{ann99} presented $UBVI$ data;  these data have also been
reanalysed by others (see Table~\ref{t:lit}). High-resolution spectra have been
obtained by \cite*{soubiran}, by \cite*{jacobson11}, and by \cite*{jacobson13} to derive
metallicity, and by \cite*{mermilliod08} addressing radial velocities. These
studies are described below (Sect.~\ref{sec:comparison}). Table~\ref{t:lit} presents a
summary of the properties derived in literature;  while all these authors agree
on an intermediate age, a moderate reddening, and a subsolar metallicity (with
one exception), an agreement on the specific values has not been reached yet.
In the present paper we analyse $BVI$ photometric data obtained at the Large
Binocular Telescope (LBT) and spectra obtained with the Nordic Optical 
Telescope  (NOT; see Sec.~2 and~3 for a description of data acquisition and
reduction), derive detailed abundances (Sec.~3), the cluster age, distance, and reddening using synthetic CMDs (Sec.~4). Discussion and summary are presented
in Secs.~5, and~6.

\section{Photometry \label{photo}}

\subsection{Observation and reduction}
\begin{figure}
\centering
\includegraphics[width=0.45\textwidth]{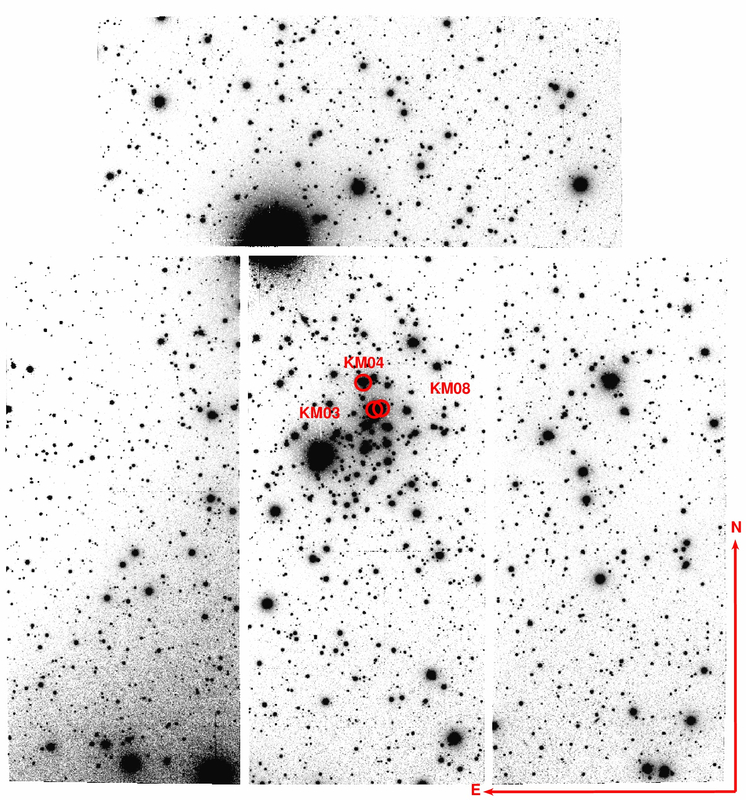} 
\caption{The field observed with LBC, corresponding to about $22\times25$
arcmin$^2$. The red circles indicate the spectroscopic targets. The image is a
composition of the CCD mosaic of the instrument obtained in the $V$-BESSEL
filter. North is up and East is left.}
\label{fov}
\end{figure}

The cluster was observed as part of the BOCCE program using LBC (Large Binocular
Camera) mounted at LBT on  Mount Graham (AZ, USA); we refer the reader to
\cite{cignoni11} and \cite{lbt2} for further details. All data were acquired during one night (Feb. 22, 2012), with seeing of about $1.6\arcsec$ and using
the Johnson-Cousins filters  ($B$ on LBC-Blue and  $V,I$ on LBC-Red). The log of
the observations is given in Table~\ref{t:log}. In Fig.~\ref{fov} we show the field of view of LBC@LBT and highlight the spectroscopic targets.

The raw LBC images were trimmed and corrected for bias and flat-field, using the
pipeline developed for LBC image pre-reduction by the Large Survey Center (LSC)
team at the Rome Astronomical Observatory.\footnote{LSC website:
http://lsc.oa-roma.inaf.it/}  The stars were detected independently on each $B$,
$V$ and $I$ image and their photometry performed using the point spread function
(PSF)-fitting code {\sc daophotii/allstar} \citep{ste_87,ste_94}.
\begin{table}
\centering
\setlength{\tabcolsep}{0.8mm}
\caption{Log of LBT and NOT observations.}
\begin{tabular}{lcll}
\hline
Instr. &UT Date          &exp.time   (s)                                                 &Note \\
\hline
LBC            &2012 Feb 22   &B: 2$\times$1.2, 3$\times$5.3, 2$\times$90.3 &seeing 1.5"-1.9"       \\
                   &                       &V: 2$\times$1.2, 3$\times$5.3, 3$\times$60.3  &seeing 1.3"-1.9"       \\
                   &                       &I: 2$\times$1.2, 5.3, 2$\times$60.3 &seeing 1.1"-1.7"       \\
\hline
FIES          &2014 Jan 25  & 3$\times$1800 & star 201728 (817) \\
                 &2014 Mar 12  & 3$\times$2100 & star 201729 (536) \\
	        &2015 Jan 05  & 3$\times$2100 &star  201735 (587)\\
\hline
\end{tabular}
\label{t:log}
\end{table}

We used the error-weighted average of the independent measurements obtained from
the various images  as the final value of the instrumental magnitudes. The
transformation of instrumental positions (in pixels) to J2000 celestial
coordinates was done\footnote{We used the code {\sc cataxcorr}, developed by
Paolo Montegriffo at the INAF - Osservatorio Astronomico di Bologna, see
http://www.bo.astro.it/$\sim$paolo/Main/CataPack.html} using, for each chip,
about 700 stars from the Sloan Digitized Sky Survey (SDSS, see www.sdss.org)
catalogue. The rms scatter of the solution was about 0.1 arcsec in both RA and
Dec. We derived the completeness level of the photometry by means of extensive
artificial stars experiments, as in our previous papers and as described in
\cite{bellazzini}.  About $10^5$ artificial stars were used to derive
photometric errors and completeness in $B$, $V$ and $I$ exposures for the
central chip. The resulting completeness ftactions are shown in
Table~\ref{compl}. For magnitudes brighter than 15, the completeness is 100\%.

\begin{table}
\centering
\caption{Completeness of our photometry (in percentage).}
\begin{tabular}{cccc}
\hline
mag & $B$ & $V$ & $I$ \\
\hline
15.0 &              & 100 $\pm$ 1 &             \\ 
15.5 & 100 $\pm$ 1  &  97 $\pm$ 2 & 100 $\pm$ 1 \\
16.0 &  97 $\pm$ 2  &  97 $\pm$ 1 &  95 $\pm$ 1 \\
16.5 &  97 $\pm$ 1  &  97 $\pm$ 1 &  95 $\pm$ 1 \\
17.0 &  98 $\pm$ 1  &  98 $\pm$ 1 &  94 $\pm$ 1 \\
17.5 &  97 $\pm$ 1  &  97 $\pm$ 1 &  92 $\pm$ 1 \\
18.0 &  98 $\pm$ 1  &  96 $\pm$ 1 &  90 $\pm$ 1 \\
18.5 &  97 $\pm$ 1  &  95 $\pm$ 1 &  84 $\pm$ 1 \\
19.0 &  96 $\pm$ 1  &  94 $\pm$ 1 &  70 $\pm$ 1 \\
19.5 &  95 $\pm$ 1  &  93 $\pm$ 1 &  27 $\pm$ 1 \\
20.0 &  95 $\pm$ 1  &  91 $\pm$ 1 &   3 $\pm$ 4 \\
20.5 &  93 $\pm$ 1  &  86 $\pm$ 1 &             \\
21.0 &  93 $\pm$ 1  &  76 $\pm$ 1 &             \\
21.5 &  90 $\pm$ 1  &  49 $\pm$ 1 &             \\
22.0 &  88 $\pm$ 1  &  16 $\pm$ 2 &             \\
22.5 &  82 $\pm$ 1  &   2 $\pm$ 5 &             \\
23.0 &  70 $\pm$ 1  &		  &             \\
23.5 &  41 $\pm$ 1  &             &             \\
24.0 &  11 $\pm$ 2  &             &             \\
24.5 &   1 $\pm$ 7  &             &             \\
\hline
\end{tabular}
\label{compl}
\end{table}

\subsection{Calibration}\label{sec:calibration}
No standard areas were observed during our observing night, thus we were forced
to calibrate the cluster by using existing  photometry. Unfortunately, neither
\citet[][who have $UBV$ images]{km91} nor \citet[][$UBVI$ images]{ann99} cover
the full FoV of LBC. The SDSS photometry cannot be used directly, because
all bright targets (i.e., giants as well as stars at the Main Sequence Turn
Off - MSTO) are saturated, so we used the SDSS to obtain calibration equations
for each of the four chips, as done in \cite{donati_tr5}. 

\begin{figure*}
\centering
\includegraphics[scale=0.42]{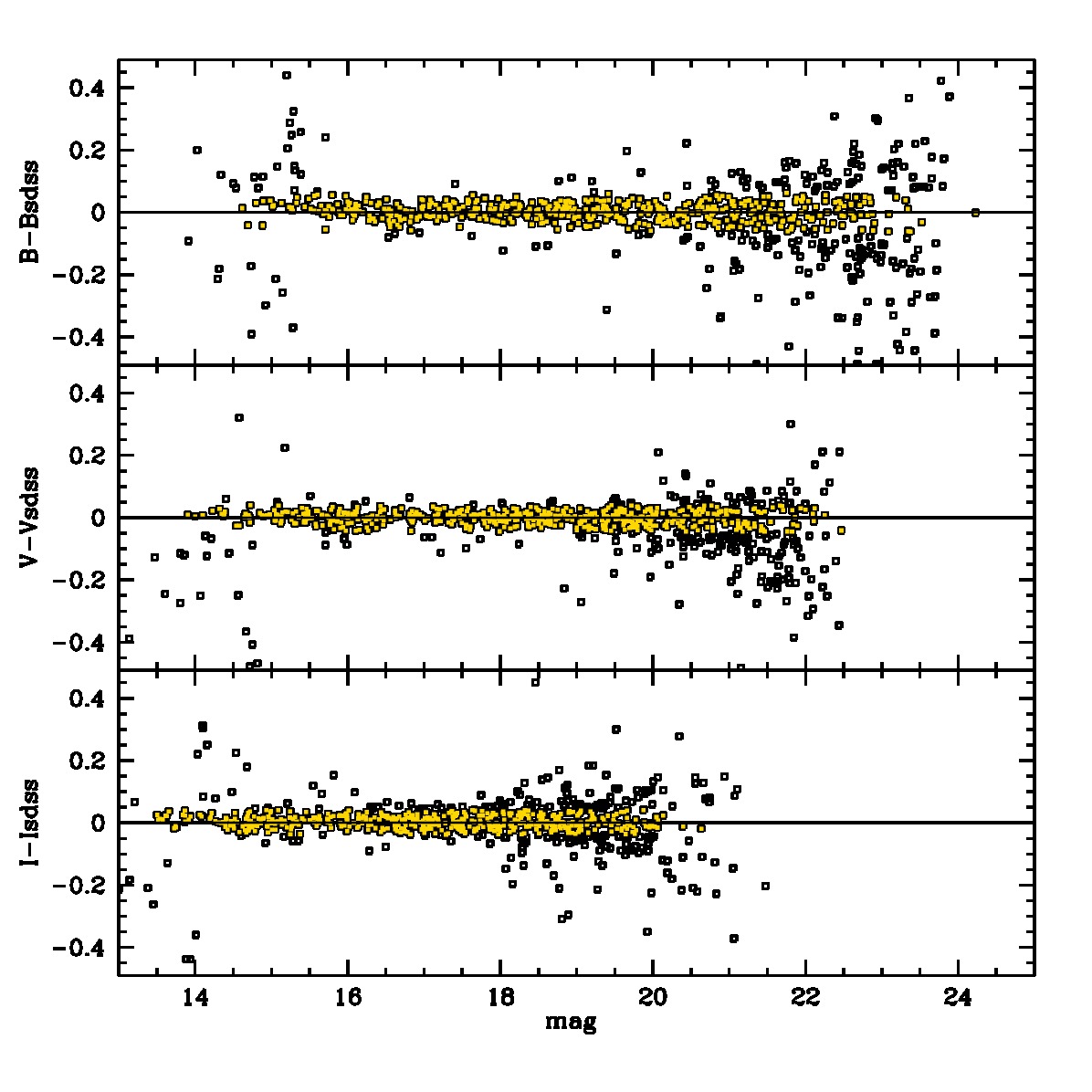} 
\includegraphics[scale=0.42]{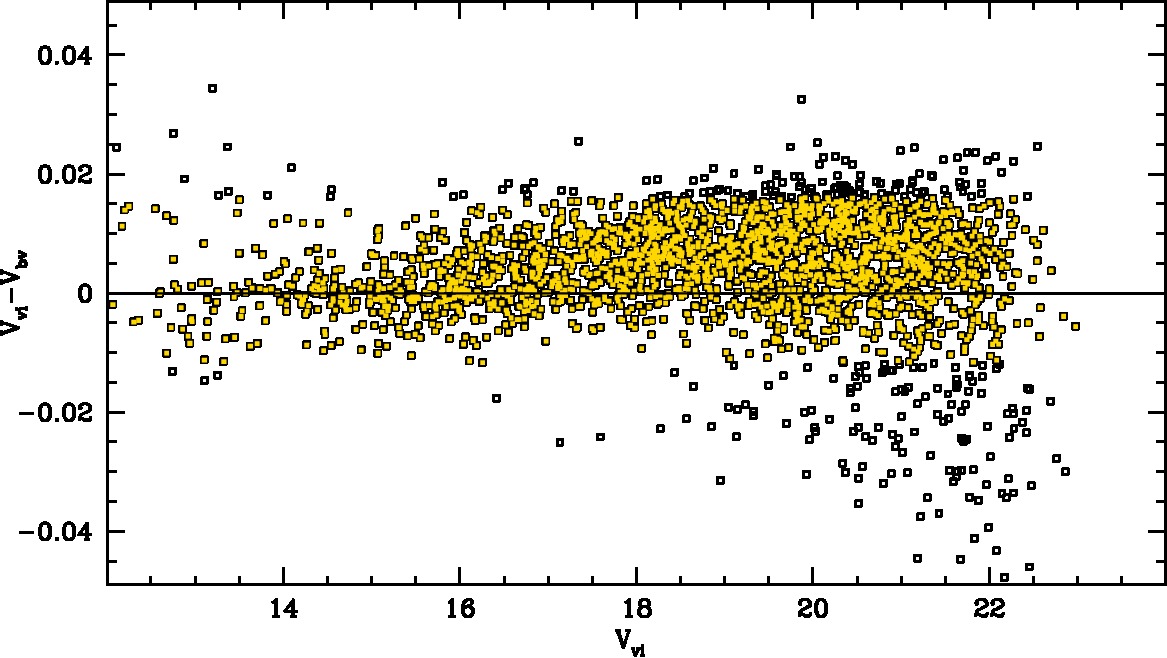} 
\caption{Left panel: Comparison of the calibrated $B$, $V$, and $I$ magnitudes
with the SDSS photometry for CCD~\#2. The yellow-filled dots are the stars used
to compute the mean difference within $1\sigma$ from the average. The
differences are consistent with 0, without trends with magnitude. The same
conclusions apply to all the other CCDs of the instrument. Right panel:
Comparison of the $V$ obtained from $b-v$ and the $V$ obtained from $v-i$.  A
slight trend with magnitude is present; however, the average difference (again
computed with the yellow filled dots within $1\sigma$ from the average) is about
0.004 mag with an rms of 0.006 mag.}
\label{fig:equcomp}
\end{figure*}

We transformed the SDSS $gri$ magnitudes to the Johnson-Cousins $BVI$
values\footnote{ For the conversion we used the equations available at
http://www.sdss.org/dr4/algorithms/sdssUBVRITransform.html\#Lupton2005}.
The calibration equations are summarised in Table~\ref{tab:equ}. They are in the
form 
$$M-m_i=zp+a\times C_i$$
for $V$ and $I$, and in the form
$$M-m_i=zp+a\times C_i+b\times C_i^2$$
for $B$, where we considered a quadratic dependence of the magnitude on the
colour. $M$ is the magnitude in the standard photometric system, $m_i$ the
instrumental magnitude, $zp$ the zero-point, $a$ describes the linear dependence
from the instrumental colour $C_i$, and $b$ describes the quadratic dependence
on the instrumental colour $C_i$. The comparisons with the original SDSS
catalogue are shown in Fig.~\ref{fig:equcomp}, where we show also the difference
between the $V$ magnitudes obtained with either the $B-V$ or the $V-I$ colours.
We deem these comparisons satisfying; in particular,  the difference between the
two $V$ magnitudes is negligible (less than 0.01 mag with a root mean square,
hereafter rms, of the same order).  

\begin{table}
 \centering
\caption{Calibration equations obtained for the four CCDs. For each equation 200 to 400 stars were used.}
\begin{tabular}{cc}
\hline
\hline
\multicolumn{2}{c}{CCD 1} \\
\hline
equation & rms \\
\hline
$B-b=4.505-0.494\times(b-v)+0.177\times(b-v)^2$ & rms 0.029 \\
$V-v=4.253-0.076\times(b-v)$ & rms 0.019 \\
$V-v=4.249-0.055\times(v-i)$ & rms 0.019 \\
$I-i=4.175-0.015\times(v-i)$ & rms 0.019 \\
\hline
\multicolumn{2}{c}{CCD 2} \\
\hline
equation & rms \\
\hline
$B-b=4.484-0.542\times(b-v)+0.232\times(b-v)^2$ & rms 0.029 \\
$V-v=4.224-0.062\times(b-v)$ & rms 0.018 \\
$V-v=4.225-0.050\times(v-i)$ & rms 0.018 \\
$I-i=4.133-0.011\times(v-i)$ & rms 0.021 \\
\hline
\multicolumn{2}{c}{CCD 3} \\
\hline
equation & rms \\
\hline
$B-b=4.489-0.554\times(b-v)+0.227\times(b-v)^2$ & rms 0.037 \\
$V-v=4.226-0.065\times(b-v)$ & rms 0.020 \\
$V-v=4.209-0.036\times(v-i)$ & rms 0.021 \\
$I-i=4.095-0.014\times(v-i)$ & rms 0.020 \\
\hline
\multicolumn{2}{c}{CCD 4} \\
\hline
equation & rms \\
\hline
$B-b=4.423-0.341\times(b-v)+0.121\times(b-v)^2$ & rms 0.039\\
$V-v=4.221-0.068\times(b-v)$ & rms 0.019 \\
$V-v=4.209-0.039\times(v-i)$ & rms 0.020 \\
$I-i=4.102-0.009\times(v-i)$ & rms 0.021 \\
\hline
\end{tabular}
\label{tab:equ}  
\end{table}

The final catalogue contains the identification, celestial coordinates,
magnitudes and errors for 4791 stars. It will be made available through the
BOCCE webpage (www.bo.astro.it/?page\_id=2632), 
WEBDA\footnote{http://webda.physics.muni.cz}, and  the Centre de Donn\`ee de
Starsbourg (CDS)\footnote{ http://cdsarc.u-strasbg.fr/viz-bin/qcat?J/MNRAS/}.

\subsection{Comparison with literature}

\begin{figure}
\includegraphics[scale=0.425]{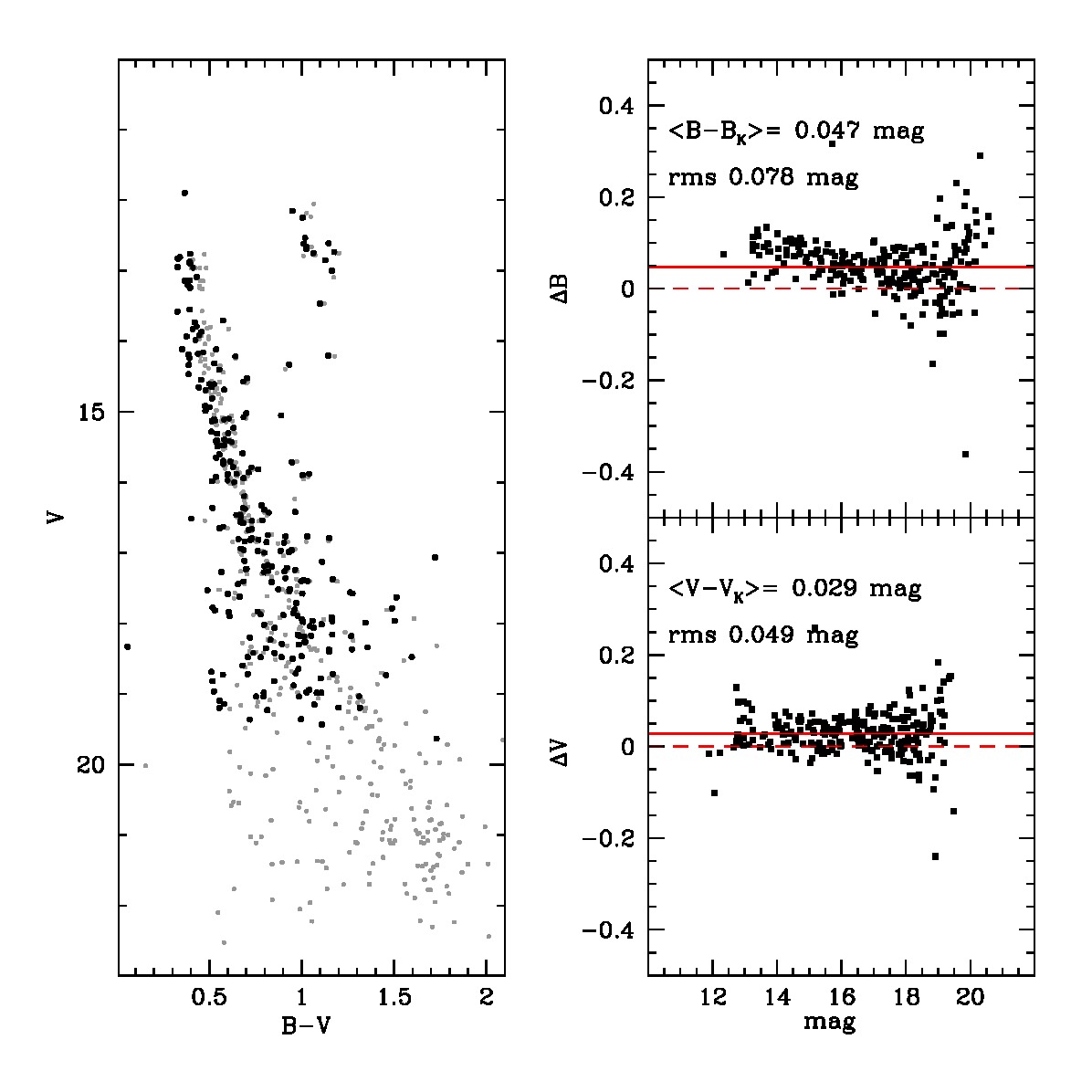}
\includegraphics[scale=0.425]{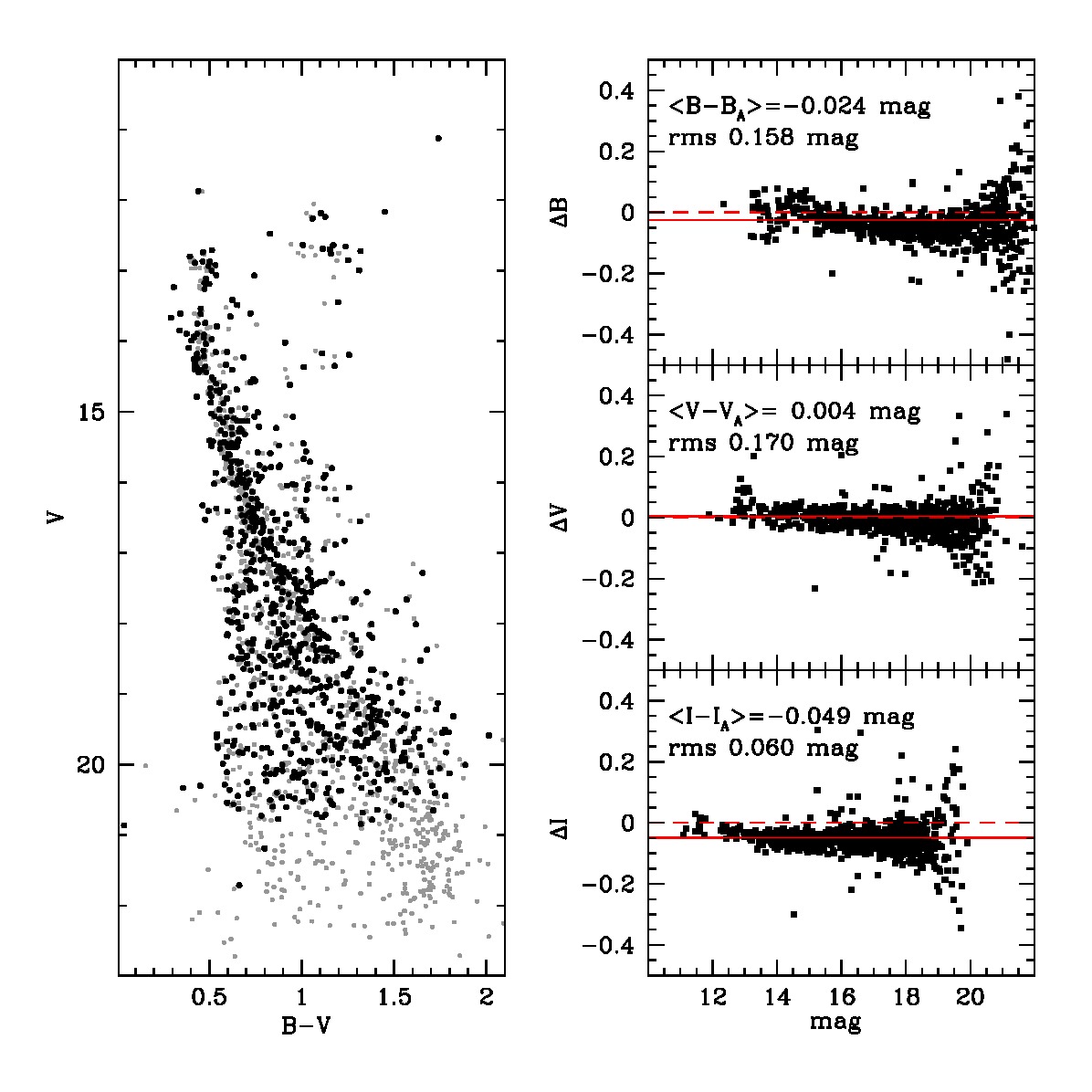}
\caption{Comparison between our photometric catalogue and the literature ones.
Upper panels: on the left the $V,B-V$ CMD from Kaluzny \& Mazur (1991) in black
and our CMD on a area of the same size in gray; on the right the comparison of
$B$ and $V$ magntidutes. The average differences and rms are indicated in the
respective panels Lower panel: the same, but with the $BVI$ photometry from Ann
et al. (1999).}
\label{figvslit}
\end{figure}

\begin{figure}
\includegraphics[scale=0.4]{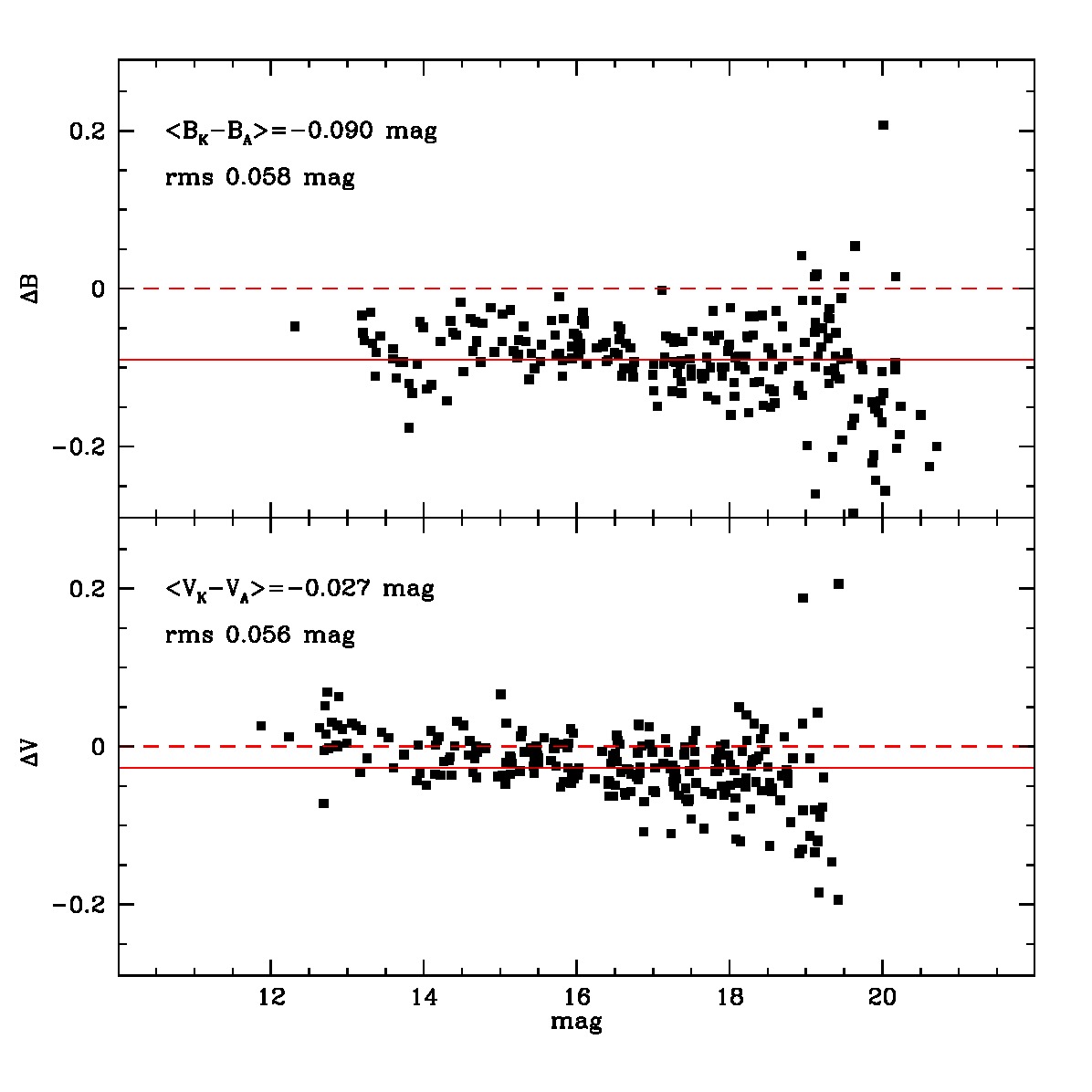}
\caption{Photometric comparison between Kaluzny \& Mazur (1991) and Ann et al.
(1999) catalogues. Upper panel: there is an offset of about $0.1$ mag in $B$,
with a complex trend with magnitude. Lower panel: there is a smaller offset in
the case of $V$, about $0.03$ mag, but also an evident dependence on magnitude.}
\label{figKvsA}
\end{figure}

As mentioned in the Introduction, only two previous studies presented CCD
photometry for this cluster \citep{km91,ann99}. We downloaded their catalogues
from the WEBDA and cross-identified stars with our photometry. In
Fig.~\ref{figvslit} we compare our photometry with theirs (upper panels for
$B,V$ and \citealt{km91}; lower panels for $B,V,I$ and \citealt{ann99}); in both
cases we reach deeper magnitudes and on a much larger field of view. As usual
when doing these comparisons, there are small offsets between different
photometries  (always smaller than $0.05$ mag, see numbers in the figure
panels). In the case of the $B$ magnitude, a colour term with both photometries
is also present.  This is absent in $V$  when using  Kaluzny and Mazur, whereas
a small trend seems to be present in $V$ or $I$ with Ann et al.'s values\footnote{Note
that the $V-I$ calibration in \cite{ann99} appears problematic; they shifted the
colours of the stars by 0.1 mag to match the same isochrone that fits the $B-V$
colour (their Fig.~7).}.

While the offsets are very small, the trend in $B$ is annoying. So we checked
that  our calibration procedure through tertiary standard stars is not affected
by significant photometric errors.  We compared directly the $B,V$ values of
\cite{km91} and \cite{ann99} and we show the results in Fig.~\ref{figKvsA}. A 
non-linear trend with magnitude is present in the $B$ filter, in addition to an
offset of almost $0.1$ mag. In the case of $V$, the offset is smaller but a
linear trend is evident. 
We are then unable to say which, if any, of the three calibrations is better;
only an independent catalogue, obtained from observations in photometric
conditions, may settle this issue.  However, the total effect on the CMD is
small, as apparent from Figure~\ref{figvslit}, and does not hamper the
determination of the cluster parameters, producing at most a small difference in
reddening and distance modulus estimation, well within the errors of the
determinations. Age is in fact mainly constrained by the difference in
magnitude  between the RC and TO stars, where offsets in $V$ cancel out.

\subsection{Centre, diameter, CMD}\label{sec:CMD}
Thanks to the precise and deep photometry of LBT and to its relatively large
FoV, we re-determined the centre of the cluster following the approach described
in \cite{donati12}. Briefly, we selected the densest region on the images by
looking for the smallest coordinates interval that contains 70\% of all the
stars. The centre is obtained as the average right ascension and declination
when the selection is iterated twice.  For a more robust estimate, several
magnitude cuts have been considered iteratively and the corresponding results
averaged. The rms on the centre coordinates is about 5$\arcsec$. We found
RA(J2000)=07:16:59.44,   
Dec(J2000)=+13:45:52.50,
to be compared with the values cited in WEBDA (RA=07:16:59, Dec=+13:45:00
referred to J2000). From the density profile it was also possible to define the
apparent diameter of the cluster. We estimated $d=9\arcmin\pm1\arcmin$ using the
radius at which the density profile flattens and reaches the background density
value. For comparison,  \cite{soubiran} found that NGC~2355 has a central
component reaching to about 7$\arcmin$, while the radius at which the surface
density drops to half the central value is 1.5$\arcmin$.    

\begin{figure*}
\includegraphics[scale=0.7]{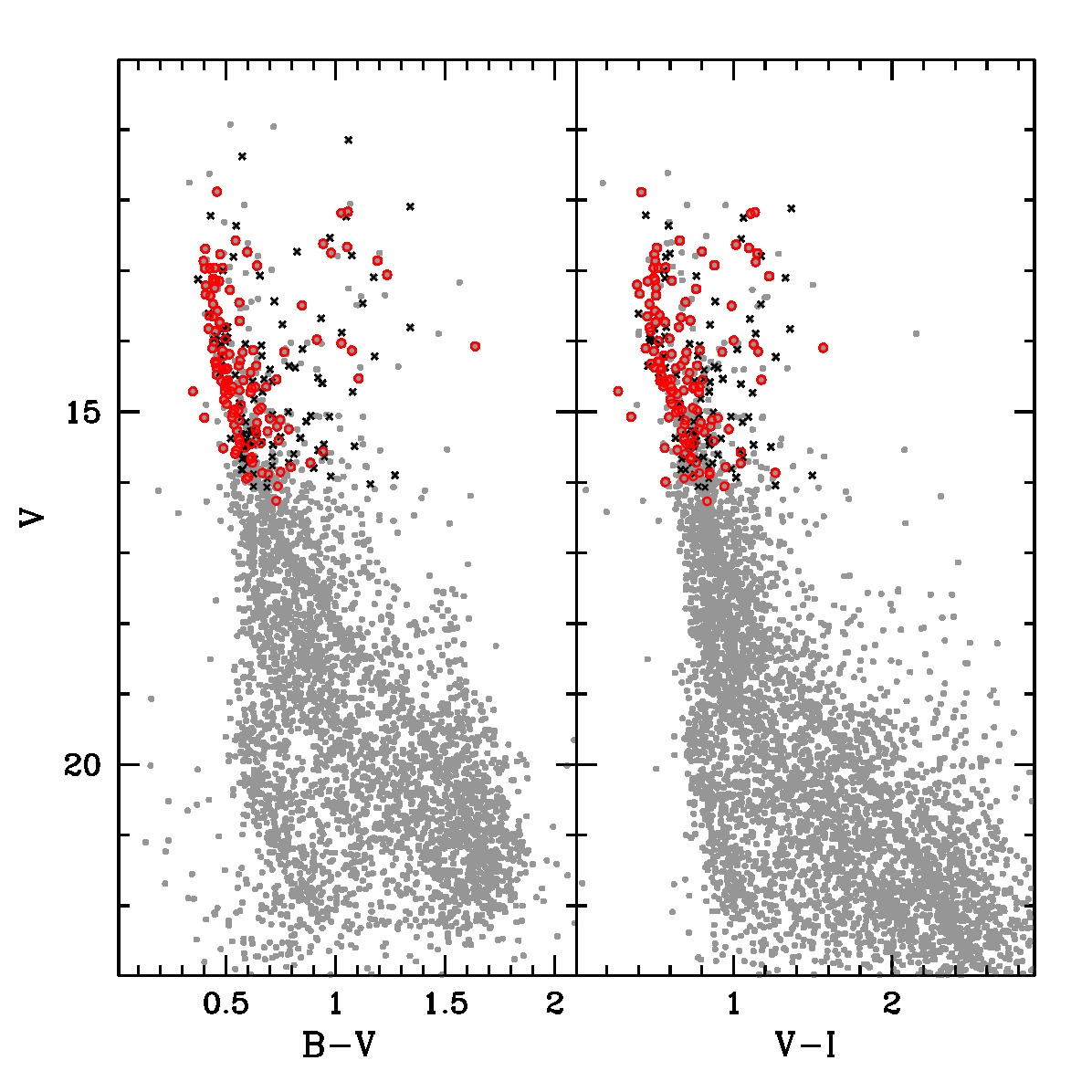}
\caption{CMDs of NGC~2355 in $V,B-V$ and $V,V-I$ for the whole LBT FoV. In red
we highlight probable members from Krone-Martins et al. 2010 (membership probability greater than 80\%), whilst black crosses are probable non-members (membership probability lower than 50\%).}
\label{figbvvi}
\end{figure*}

Membership probability based on proper motions were published by 
\cite{kronemartins} for almost 400 stars in the field of NGC~2355,  based on 
the PM2000 catalogue \citep{pm2000}, complete to $V\sim15$ mag and with a
limiting magnitude $V\sim16$ mag. Unfortunately, we cannot use these magnitudes
to solve the problem of the differences among photometries, because they are not
on the standard Johnson system (see \citealt{bordeaux}). However, this
information is useful to better define the cluster evolutionary sequences.
\cite{kronemartins} found 213 probable member stars; we cross-identified their
catalogue with our photometry and use their membership determination and
membership from RV (both analyses are in quite good agreement: 18 stars out of
23 in common between them are defined cluster members) to distinguish the
cluster footprint (see Fig.~\ref{figbvvi}) and to derive the cluster parameters
(see Sec.~\ref{sec:synth}). 

NGC~2355 is not a very popolous cluster, as shown by the CMD in
Figure~\ref{figbvvi}.  However, the cluster MS is well distinguishable with
respect to the field contamination. Field stars possess  a complex pattern in
the CMD, with at least three major components: a faint and blue MS, signature of
a much more distant, older population, most likely the thick disc;  an
intermediate MS with stars defining an almost vertical sequence;  and the
vertical locus of M dwarfs, redder than the cluster MS, evident particularly in
the $V,B-V$ CMD.  A study of the Galactic field population is beyond the scope
of this paper, but it can be a side product of the BOCCE project, especially
when large FoV are employed \citep[see][]{cignoni08,cignoni11}. 

Field interlopers, though mainly located below the cluster MS, complicate the
interpretation of the upper MS, especially the identification of the red-hook
(RH, the reddest point on the MS before the overall contraction evolutionary
phase) and the giant phase which are poorly populated. Taking the membership
into account, thanks to the radial velocity measures available in literature
(see references in the Introduction) and the proper motion analysis \citep[see][]{kronemartins}, it is possible to
define the sequences of the cluster with more confidence. However, as apparent
from Fig.~\ref{figbvvi}, the identification of the RH is not straightforward.
Therefore we decided to conservatively identify only the MS termination point
(MSTP), as the brightest luminosity level reached by MS stars. 

\begin{figure*}
\includegraphics[scale=0.7]{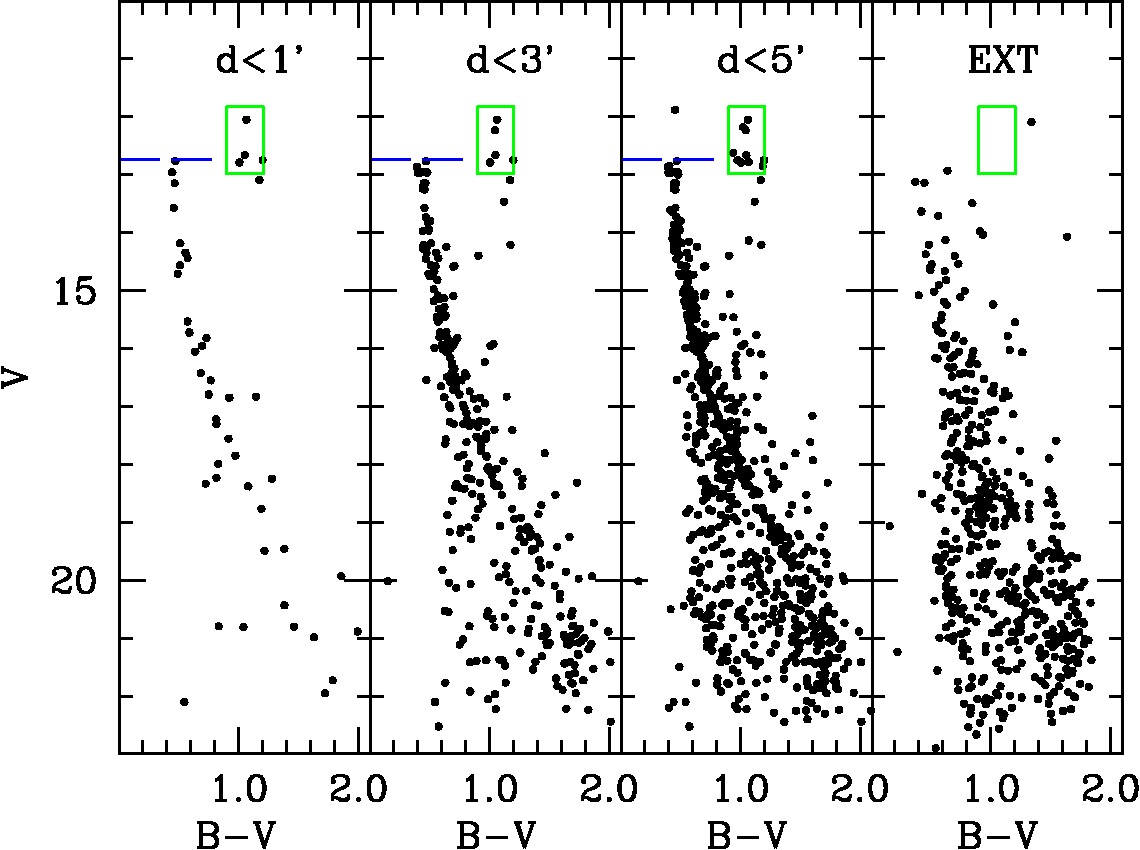}
\caption{CMDs for regions within different distances from the cluster centre,
compared with an external field of a circular area of $5\arcmin$ radius. We
highlight the  MSTP and RC. For a better comparison, the RC box is shown also in
the CMD of the external field.} 
\label{figrad}
\end{figure*}

In Fig.~\ref{figrad} we show the CMDs of NGC~2355 for regions within different
distances from the cluster centre ($1\arcmin$, $3\arcmin$, and $5\arcmin$) in
comparison with an external area of $5\arcmin$ radius. We identify a small gap
at about $V\sim13.4$ mag on the MS, also found by both \cite{km91} and
\cite{ann99}; the occurrence of similar gaps is not uncommon, in open clusters
(e.g., NGC~6134, \citealt{ahumada13}) and in principle can be related to the overall contraction phase, however the following analysis with synthetic CMDs (see Sec.~\ref{sec:synth}) does not give a firm confirmation of this interpretation.

We used the radial plots and the membership probabilities to identify the
evolutionary features of the cluster employed in the following analysis (see
Sec.~\ref{sec:synth}), and we identified:
\begin{itemize}
  \item the RC and giants at $V\sim 12.5$, $(B-V)\sim 1$ mag;
  \item the MSTP at $V\sim 13$ mag;
  \item and the MS extending down to $V\sim 22$ mag.
\end{itemize}

\section{Spectroscopy \label{spectro}}
As already mentioned in the Introduction, several stars in NGC~2355 have been
observed with high-resolution spectrocopy,  but our study adds new important
information. In fact, M08 only derived radial velocity (RV); S00 obtained rather
low signal-to-noise ratio (S/N) spectra,  useful for RV, atmospheric parameters,
and metallicity, but not suitable for detailed chemical analysis. J11 have
spectra covering only a limited wavelength range, so that only a few elements could
be measured; finally, \cite{jf} observed only three stars. Thus  our work
doubles the number of stars for which abundances of elements forged by different
nucleosynthetic chains can be determined in this cluster.

\subsection{Observation and reduction}

We selected stars on the RC of NGC~2355 as possible targets, using the
information from S00 and M08 to observe only members (see Table~\ref{altrerv}). We obtained
high-resolution spectra of three RC stars  using FIES (FIbre-fed Echelle
Spectrograph) at the NOT  (see \citealt{fies} for instrumentation details). FIES
covers the range $\lambda\lambda=3700-7300$ \AA \ without gaps. We used the
medium-resolution fibre bundle ($R \equiv \lambda/\Delta\lambda=46000$) and a
$2\times2$ binning to increase the S/N.  

\begin{figure}
\centering
\includegraphics[scale=0.8]{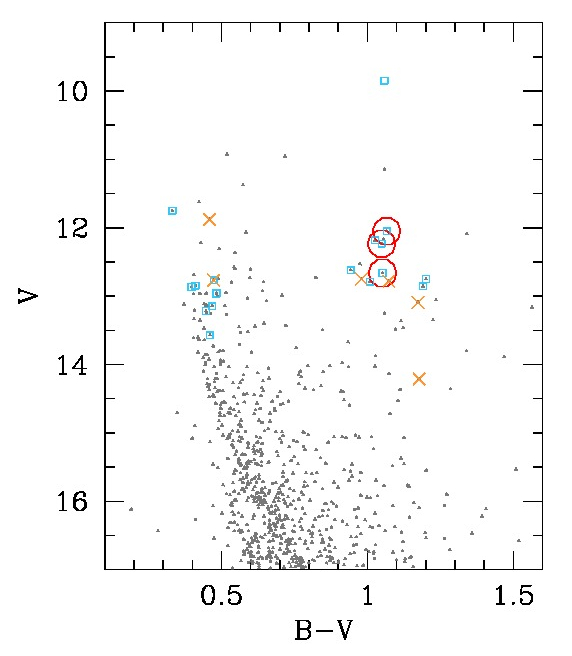}
\caption{CMD of NGC~2355 with the three stars observed with NOT@FIES indicated by large red circles. Light blue squares indicate member stars and orange crosses non members (from \citealt{soubiran,mermilliod08,jacobson11}) in common with our photometry.}
\label{cmdrv}
\end{figure}

Information on the observed stars is provided in Table~\ref{spe1} and their
position in the CMD  is indicated in Fig.~\ref{cmdrv}.  Star \#817 was observed
in visitor mode in January 2014, while stars \#536, 587 were subsequently
observed in service mode (see Table~\ref{t:log}). For the second star, no
wavelength calibration frames were taken in the same night; we had to use  the
January lamps and this resulted in a lower precision in the zero point of the
calibration. However,  this is irrelevant to chemical analysis, but it results
in a less precise RV measurement.

\begin{table}
\centering
\caption{Stars in common with S00, M08, J11 and their status.}
\setlength{\tabcolsep}{1.25mm}
\begin{tabular}{lclrrr}
\hline
ID        &WEBDA &ID(S00)     &S00     &M08   &J11 \\
\hline
\multicolumn{6}{c}{High-res spectra and photometry}\\
   201728 &817  &       km03  & yes  &   M  & ... \\
   201729 &536  &	km04  & yes  &   M  & ... \\
   201735 &587  &	km08  & yes  &   M  & ... \\
\multicolumn{6}{c}{Only photometry}\\
   201712 &335  &	km19  & yes  &  ... & ... \\	 
   201715 &662  &	km22  & yes  &  ... & ... \\
   201720 &441  &	km21  & yes  &  ... & ... \\
   201726 &668  &GSC77500538  & yes  &  ... & M   \\
   201731 &276  &	km02  &  no  &  ... & ... \\
   201732 &563  &	km10  & yes  &   M  & M   \\
   201734 &832  &GSC77501264  & yes  &  ... & ... \\
   201736 &736  &	km07  &  no  &  NM  & NM  \\
   201737 &734  &	...   & ...  &  ... & NM  \\
   201738 &592  &	km09  & yes  &   M  & M   \\
   201739 &377  &	km15  & yes  &  ... & M   \\
   201740 &599  &	km20  &  no  &  SB  & ... \\
   201742 &734  &	km12  & yes  &  ... & ... \\
   201744 &382  &	km13  & yes  &  ... & ... \\
   201745 &296  &	km14  & yes  &  ... & ... \\
   201752 &360  &	...   & ...  &  ... & NM  \\
   201754 &472  &	km27  & yes  &  ... & ... \\
   301356 &144  &GSC77501198  & yes  &  ... & M   \\
\hline
\end{tabular}
\begin{list}{}{}
\item[] KM is the identifier in \cite{km91}
\item[] M09, J11 use the WEBDA identifier.
\item[] yes,M : member stars ; NM : not a member ; SB: binary
\end{list}
\label{altrerv}
\end{table}
%
%

Star \#817 was observed with three 1800s-long exposures while stars \#536 and \#587 had three 2100s-long exposures; they were reduced using the dedicated FIES software ({\sc FIEStool}), which takes care of all standars steps, from bias subtraction to wavelength calibration. We cleaned the telluric contamination in a small region near the [O {\sc i}] line and measured the RV on the final 1-d spectra using {\sc iraf}\footnote{
{\sc iraf} is the Image Reduction and Analysis Facility, distributed by the National Optical Astronomy Observatory, which is operated by the Association of Universities for Research in Astronomy (AURA) under cooperative agreement with the National Science Foundation.} 
routines and {\sc daospec} \citep{daospec}.  The individual spectra were then shifted and co-added.

\begin{table*}
\setlength{\tabcolsep}{1.25mm}
\caption{Data for the three stars with high-resolution spectra.}
\begin{tabular}{ccccccccccccc}
\hline
 ID   & WEBDA & other & RA	    & Dec	 & B	 & V	 & S/N & RV	&T$_{\rm eff}$ &$\log$g &$v_t$ &[Fe/H] \\
      &       &	&(J2000)     &(J2000)	 &	 &	 &     &km/s	&	   &	    &	   &	   \\
\hline
201728& 817    & KM03	&109.2414551 &+13.7734194 &13.116 &12.050  & 80 & 35.57  &5080 & 2.64 &1.16 &-0.01 \\	  
201729& 536    & KM04	&109.2476883 &+13.7881250 &13.283 &12.234  & 90 & 39$^a$ &4998 & 2.41 &1.19 &-0.04 \\	  
201735& 587    & KM08	&109.2376022 &+13.7740917 &13.714 &12.662  & 110 & 35.44  &5110 & 2.88 &1.13 &-0.15 \\	  
\hline
\end{tabular}
\begin{list}{}{}
\item[] Notes - other=\cite{km91}. 
\item[] $^a$ The RV for star 536 is uncertain (see text).
\end{list}
\label{spe1}
\end{table*}

\subsection{Atmospheric parameters and iron abundances}

These three spectra were analysed as in other BOCCE clusters and we give here
only a short description of the procedure; for more details see
\cite{bragaglia01,bragaglia06,carretta04,carretta05}. Equivalent widths (EWs)
were measured employing an updated version of the {\sc rosa} spectrum analysis
package \citep{rosa}. We restricted to the 5500-7000 \AA \ spectral range for Fe
lines to minimize problems of line crowding and difficult continuum tracing
blueward of this region and of contamination by telluric lines redward. We
employed the entire spectrum for other elements. Sources of oscillator strengths
and atomic parameters are the same as in \cite{gratton03}.

We used as intial guesses for effective temperature (T$_{\rm eff}$) and gravity ($\log g$) the values based on
photometric data, the \cite{alonso} relations,  distance, and
reddening. We then derived the final values for T$_{\rm eff}$ and $\log g$ from the spectra  using the excitation and ionisation equilibria for iron, respectively. The microturbulent velocity ($v_t$) was derived assuming the relation between $\log g$ and $v_t$ given in \cite{carretta04}, i.e., $v_t = 1.5 \times \log g$. These parameters, along with iron abundances, are reported in Table~\ref{spe1}. 

The iron abundances derived from EWs were checked using synthetic spectra of about 30 selected iron lines \citep[see][for a description of these lines]{carretta04}. The average differences between iron abundance based on EW and on synthesis is 0.1 dex, without systematic trends. We are then confident to have achieved the accuracy in continuum tracing and EW measurements possible with these spectra.
The average values we found for the three stars are [Fe/H]{\sc
i}$=-0.064\pm0.044$ (rms=0.076) dex, and [Fe/H]{\sc ii}$=-0.066\pm0.043$
(rms=0.075) dex. The very good agreement between the values for two ionisation states (found also for Ti and Cr, see later) is a further confirmation of the surface gravities.

We derived errors in EWs, temperature, gravity, model metallicity, and
microturbulent velocity taking into account that they are not independent.
Errors were estimated as in Carretta et al. (2004) where a detailed description
is given. They comprise a) a random part different from star-to-star (due, for
example, to the different S/N ratios) that represents the internal error; and b)
a systematic part (due, for example, to uncertainties in the adopted oscillator
strengths, blends not considered, etc.). Random errors have been found to be of
25 K in T$_{\rm eff}$, 0.07 dex in $\log g$, and 0.17 km~s$^{-1}$ in $v_t$. In
Table~\ref{sens} we present the sensitivity of the derived abundance ratios to
the variations in atmospheric parameters and to errors in EWs, obtained by
repeating our abundance analysis by changing only one parameter at the time. The
amount of the changes is shown in the first line of the header, the random error
used to determine the resulting  abundance changes is given in the second line,
and the corresponding variations are given for all elements separately for each
parameter and as a total  (internal) variation.

\begin{table*}
\centering
\caption[]{Sensitivities of abundance ratios to variations in the atmospheric
parameters and to errors in the equivalent widths (or fit, for specteum
synthesis), and errors in abundances for stars of NGC~2355.}
\begin{tabular}{lrrrrrrrr}
\hline
Element     & Average   & T$_{\rm eff}$ & $\log g$ & [A/H]   & $v_t$    & EWs     & Total   & Note\\
            & n. lines  &      (K)      &  (dex)   & (dex)   &kms$^{-1}$& (dex)   &Internal &\\
\hline        
Variation&              &  50           &   0.20   &  0.10   &  0.10    &         &         &\\
Internal &              &  25           &   0.07   &  0.07   &  0.17    & 0.06    &         &\\
\hline
$[$Fe/H$]${\sc  i}& 108 &    +0.038	& $-$0.001 &  +0.004 & $-$0.040 & 0.005  & 0.071    &EW\\
$[$Fe/H$]${\sc ii}&  14 &  $-$0.032	&   +0.095 &  +0.031 & $-$0.033 & 0.015  & 0.073    &EW\\
$[$O/Fe$]${\sc  i}&   2 &  $-$0.031	&   +0.089 &  +0.036 &   +0.039 & 0.040  & 0.089    &EW\\
$[$Na/Fe$]${\sc i}&   4 &  $-$0.004	& $-$0.022 &  +0.002 &   +0.014 & 0.029  & 0.038    &EW\\
$[$Mg/Fe$]${\sc i}&   3 &  $-$0.013	& $-$0.008 &$-$0.005 &   +0.022 & 0.033  & 0.050    &EW\\
$[$Al/Fe$]${\sc i}&   2 &  $-$0.008	& $-$0.007 &$-$0.006 &   +0.029 & 0.040  & 0.064    &EW\\
$[$Si/Fe$]${\sc i}&  13 &  $-$0.035	&   +0.050 &  +0.012 &   +0.055 & 0.016  & 0.099    &EW\\
$[$Ca/Fe$]${\sc i}&  14 &    +0.010	& $-$0.028 &$-$0.007 & $-$0.010 & 0.015  & 0.026    &EW\\
$[$Sc/Fe$]${\sc ii}&  6 &    +0.026	& $-$0.009 &  +0.000 & $-$0.011 & 0.023  & 0.033    &EW\\
$[$Ti/Fe$]${\sc i}&   9 &    +0.025	& $-$0.004 &$-$0.007 &   +0.013 & 0.019  & 0.032    &EW\\
$[$Ti/Fe$]${\sc ii}&  1 &    +0.022	& $-$0.005 &  +0.003 &   +0.027 & 0.057  & 0.074    &EW\\
$[$Cr/Fe$]${\sc i}&  10 &    +0.005	& $-$0.006 &$-$0.004 &   +0.018 & 0.018  & 0.036    &EW\\
$[$Cr/Fe$]${\sc ii}&  2 &    +0.003	& $-$0.008 &$-$0.006 &   +0.020 & 0.040  & 0.053    &EW\\
$[$Mn/Fe$]${\sc i}&   4 &    +0.016	& $-$0.036 &  +0.002 & $-$0.029 & 0.029  & 0.059    &EW\\
$[$Cu/Fe$]${\sc i}&   1 &    +0.050     &   +0.040 &  +0.040 &   +0.030 & 0.050  & 0.095    &SS\\
$[$Ni/Fe$]${\sc i}&  35 &  $-$0.008	&   +0.020 &  +0.006 &   +0.008 & 0.010  & 0.019    &EW\\
$[$Zn/Fe$]${\sc i}&   1 &  $-$0.049	&   +0.049 &  +0.021 & $-$0.005 & 0.057  & 0.067    &EW\\
$[$Y/Fe$]${\sc ii}&   3 &    +0.050     &   +0.030 &  +0.000 &   +0.090 & 0.035  & 0.111    &SS\\
$[$Zr/Fe$]${\sc i}&   3 &    +0.044	& $-$0.002 &$-$0.002 &   +0.039 & 0.033  & 0.077    &EW\\
$[$Zr/Fe$]${\sc ii}&  1 &    +0.029	& $-$0.007 &  +0.003 &   +0.017 & 0.057  & 0.066    &EW\\
$[$Ba/Fe$]${\sc ii}&  4 &    +0.030     &   +0.050 &  +0.000 &   +0.080 & 0.006  & 0.099    &SS\\
$[$La/Fe$]${\sc ii}&  3 &    +0.040     &   +0.050 &  +0.020 &   +0.050 & 0.021  & 0.085    &SS\\
$[$Ce/Fe$]${\sc ii}&  1 &    +0.038	& $-$0.008 &  +0.006 &   +0.014 & 0.057  & 0.065    &EW\\
$[$Pr/Fe$]${\sc ii}&  3 &    +0.039	& $-$0.009 &  +0.007 &   +0.026 & 0.033  & 0.059    &EW\\
$[$Nd/Fe$]${\sc ii}&  2 &    +0.030     &   +0.040 &  +0.020 &   +0.050 & 0.028  & 0.079    &SS\\ 
$[$Sm/Fe$]${\sc ii}&  2 &    +0.045	& $-$0.007 &  +0.010 &   +0.011 & 0.040  & 0.050    &EW\\
$[$Eu/Fe$]${\sc ii}&  2 &    +0.050     &   +0.050 &  +0.000 &   +0.060 & 0.070  & 0.116    &SS\\

\hline
\end{tabular}
\label{sens}
\end{table*}

\begin{table*}
\centering
\caption{Abundances for the three stars; we indicate when spectrum synthesis
(SS) or equivalent widths (EW) were used.}
\begin{tabular}{lrrrrrrrrrrrrc}
\hline
 &\multicolumn{4}{l}{201729}&\multicolumn{4}{l}{201728}&\multicolumn{4}{l}{201735} &Note\\
element   	&nr &logn  &[X/Fe] &rms  &nr &logn  &[X/Fe] &rms &nr &logn  &[X/Fe] &rms&    \\ 
\hline
${\rm [Fe/H]}${\sc i}    &108 & 7.500 &-0.040 &0.102 &107 & 7.536 &-0.004 &0.094 &109 & 7.391 &-0.149 &0.070& EW\\
${\rm [Fe/H]}${\sc ii}   & 14 & 7.443 &-0.047 &0.080 & 14 & 7.488 &-0.002 &0.094 & 13 & 7.341 &-0.149 &0.067& EW\\
${\rm [O/Fe]}$      	 &  2 & 8.524 &-0.226 &0.075 &  2 & 8.380 &-0.406 &0.010 &  1 & 8.558 &-0.083 &     & EW\\
${\rm [Na/Fe]}$     	 &  4 & 6.353 & 0.183 &0.042 &  4 & 6.383 & 0.177 &0.074 &  4 & 6.200 & 0.139 &0.138& EW\\
${\rm [Mg/Fe]}$     	 &  3 & 7.456 & 0.066 &0.082 &  3 & 7.425 &-0.001 &0.026 &  3 & 7.410 & 0.129 &0.073& EW\\
${\rm [Al/Fe]}$     	 &  2 & 6.114 &-0.076 &0.091 &  2 & 6.249 & 0.023 &0.093 &  2 & 6.123 & 0.042 &0.117& EW\\
${\rm [Si/Fe]}$     	 & 14 & 7.549 & 0.059 &0.097 & 14 & 7.537 & 0.011 &0.119 & 12 & 7.415 & 0.034 &0.073& EW\\
${\rm [Ca/Fe]}$     	 & 14 & 6.386 & 0.156 &0.091 & 16 & 6.372 & 0.106 &0.114 & 13 & 6.216 & 0.095 &0.123& EW\\
${\rm [Ti/Fe]}${\sc i}   &  8 & 4.850 &-0.110 &0.083 &  9 & 4.953 &-0.043 &0.061 &  9 & 4.815 &-0.036 &0.061& EW\\
${\rm [Ti/Fe]}${\sc ii}  &  1 & 4.904 &-0.119 &      &  1 & 4.947 &-0.121 &	 &  2 & 4.825 &-0.096 &0.030& EW\\
${\rm [Sc/Fe]}${\sc ii}  &  5 & 3.069 &-0.014 &0.111 &  7 & 3.071 &-0.057 &0.069 &  5 & 2.949 &-0.032 &0.149& EW\\
${\rm [Cr/Fe]}${\sc i}   &  9 & 5.571 &-0.059 &0.129 & 10 & 5.620 &-0.046 &0.121 & 10 & 5.484 &-0.037 &0.124& EW\\
${\rm [Cr/Fe]}${\sc ii}  &  1 & 5.576 &-0.087 &      &  1 & 5.663 &-0.045 &	 &  3 & 5.534 &-0.027 &0.011& EW\\
${\rm [Mn/Fe]}$    	 &  3 & 5.444 & 0.144 &0.057 &  3 & 5.391 & 0.055 &0.039 &  6 & 5.224 & 0.033 &0.141& EW\\
${\rm [Cu/Fe]}$          &  1 &       &-0.25  &      &  1 &	  &-0.30  &	 &  1 &       &-0.20  &     & SS\\
${\rm [Ni/Fe]}$    	 & 34 & 6.125 &-0.115 &0.068 & 34 & 6.183 &-0.093 &0.089 & 36 & 6.080 &-0.051 &0.070& EW\\
${\rm [Zn/Fe]}$    	 &  1 & 4.491 &-0.059 &      &  2 & 4.495 &-0.091 &0.029 &  1 & 4.487 & 0.046 &     & EW\\
${\rm [Y/Fe]}${\sc ii}   &  2 &       &-0.05  &0.05  &  2 &	  &-0.18  &0.13  &  2 &       &-0.10  &0.10 & SS\\
${\rm [Zr/Fe]}${\sc i}   &  3 & 2.363 &-0.190 &0.034 &  3 & 2.390 &-0.208 &0.037 &  3 & 2.319 &-0.132 &0.022& EW\\
${\rm [Zr/Fe]}${\sc ii}  &  1 & 2.363 &-0.190 &      &  1 & 2.378 &-0.220 &	 &  1 & 2.329 &-0.122 &     & EW\\
${\rm [Ba/Fe]}${\sc ii}  &  3 &       & 0.20  &0.01  &  3 &	  &-0.03  &0.03  &  3 &       &-0.03  &0.04 & SS\\
${\rm [La/Fe]}${\sc ii}  &  2 &       & 0.12  &0.03  &  2 &	  & 0.04  &0.04  &  2 &       & 0.20  &0.01 & SS\\
${\rm [Ce/Fe]}${\sc ii}  &  1 & 1.701 & 0.118 &      &  1 & 1.737 & 0.109 &	 &  1 & 1.731 & 0.250 &     & EW\\
${\rm [Pr/Fe]}${\sc ii}  &  3 & 0.561 &-0.102 &0.088 &  3 & 0.536 &-0.172 &0.130 &  4 & 0.739 & 0.178 &0.138& EW\\
${\rm [Nd/Fe]}${\sc ii}  &  2 &       & 0.03  & 0.04 &  2 &	  &-0.03  &0.04  &  2 &       & 0.16  &0.05 & SS\\
${\rm [Eu/Fe]}${\sc ii}  &  1 &       &-0.40  &      &  1 &	  &-0.40  &	 &  1 &       &-0.25  &     & SS\\
\hline
\end{tabular}
\label{abu}
\end{table*}

\subsection{Other elements}

We derived abundances for O, the light elements Na and Al,  the $\alpha$-process
elements Mg, Si, Ca, Ti (from Ti~{\sc i} and {\sc ii} lines), the Fe-group
elements Sc, Cr (from neutral and singly-ionised features), Mn, Ni, Cu, Zn, and
for the neutron-capture elements Y, Zr, Ba, La, Ce, Pr, Nd, Eu. The abundance
ratios for the three stars are given in Table~\ref{abu}, together with the
number of lines used, the abundance by number, and the rms.

As done in previous works of the series, most of the abundances were derived using EWs; we took hyperfine structure into
consideration for Sc and Mn and the abundance of Na was corrected for departures
from local thermodynamic equilibrium (LTE) according to \cite{gratton99}. For O,
we checked the abundance also with spectrum synthesis.

Abundance analyses for Cu and the neutron-capture elements Y, Ba, La, Nd, and Eu
have been carried out via spectral synthesis calculations using {\sc MOOG}
(\citealt{sneden73}, 2014 version) and the Kurucz (\citeyear{kurucz93}) grid of
model atmospheres, with solar-scaled composition and convective overshooting. 
This is different from what we did in past papers so a few details on the features used are given below.

Copper abundances were determined from the 5782 \AA~line, taking into account
the hyperfine structure (from \citealt{steffen85}) and adopting isotopic ratios
of 69\% and 31\% for $^{63}$Cu and $^{65}$Cu, respectively.

We employed a single-line treatment for Y~{\sc ii} lines at 4884~\AA, 4900~\AA,
and 5729~\AA.  For the first two spectral features we adopted the same line
lists as in \cite{dorazi13},  whereas atomic parameters ($\chi$=1.84 eV,
log$gf$=$-$1.12) for the line at 5728 \AA~come from \cite{thygesen14}. McWilliam
et al. (2013) published hyperfine structure splitting for the last spectral line
(with a total log$gf$ of $-$0.99).  However, we checked that the difference in
the resulting abundance with respect to a single-line treatment is relatively
small (0.05 dex), which is well within our observational uncertainties. This is
not surprising, because, although the main Y isotope ($^{89}$Y) has an odd mass
number, the level splitting is negligible, given the small spin and magnetic
momentum of the yttrium nucleus. Thus, for sake of consistency with the other
lines under consideration in the present study we did not include hyperfine
splittings. The standard deviation (see Table~\ref{mean}) of the average Y abundances
ensures us that our assumption is quite reasonable.

Ba abundances were inferred by analysing single-ionised features at 5853 \AA, 6141 \AA~and 6496 \AA, adopting hyperfine and isotopic splitting following 
\cite{mcwilliam98} and assuming solar mixtures for the Ba isotopes, namely 81\% for ($^{134}$Ba+$^{136}$Ba+$^{138}$Ba) and 19\% for ($^{135}$Ba+$^{137}$Ba), as in our previous works (see e.g., \citealt{dorazi12}). 

The log$gf$ for all the components for La~{\sc ii} lines at 4748 \AA, 4804 \AA, and 6390 \AA~and  Eu~{\sc ii} at 6645 \AA~were retrieved from \cite{lawler01} and \cite{lawler01b}, respectively. In the latter case isotopic ratios of 48\% and 52\% for $^{151}$Eu and $^{153}$Eu, respectively, were chosen. 

Finally, Nd abundances come from Nd~{\sc ii} spectral features at 5740~\AA and 5811~\AA, with atomic parameters provided by \cite{denhartog03}.

In Table~\ref{mean} we give the average abundance ratios for three elements measured in the three stars of NGC~2355, together with the error and rms.
We also give the solar reference values adopted in the analysis.

\begin{table}
\centering
\caption{Cluster averages and solar reference abundances}
\begin{tabular}{lrrrr}
\hline
element &mean &$\pm$ &rms & Sun  \\
\hline
${\rm [Fe/H]}${\sc i}     &-0.064 &0.044 &0.076  & 7.54\\
${\rm [Fe/H]}${\sc ii}    &-0.066 &0.043 &0.075  & 7.49\\ 
${\rm [O/Fe]}$      	  &-0.238 &0.094 &0.162  & 8.79\\ 
${\rm [Na/Fe]}$     	  & 0.166 &0.014 &0.024  & 6.21\\ 
${\rm [Mg/Fe]}$     	  & 0.065 &0.038 &0.065  & 7.43\\ 
${\rm [Al/Fe]}$     	  &-0.004 &0.036 &0.063  & 6.23\\ 
${\rm [Si/Fe]}$     	  & 0.035 &0.014 &0.024  & 7.53\\ 
${\rm [Ca/Fe]}$     	  & 0.119 &0.019 &0.033  & 6.27\\ 
${\rm [Ti/Fe]}${\sc i}    &-0.063 &0.024 &0.041  & 5.00\\ 
${\rm [Ti/Fe]}${\sc ii}   &-0.112 &0.008 &0.014  & 5.07\\ 
${\rm [Sc/Fe]}${\sc ii}   &-0.034 &0.013 &0.022  & 3.13\\ 
${\rm [Cr/Fe]}${\sc i}    &-0.047 &0.006 &0.011  & 5.67\\
${\rm [Cr/Fe]}${\sc ii}   &-0.053 &0.018 &0.031  & 5.71\\ 
${\rm [Mn/Fe]}$    	  & 0.077 &0.034 &0.059  & 5.34\\ 
${\rm [Cu/Fe]}$    	  &-0.250 &0.029 &0.050  & 4.19\\
${\rm [Ni/Fe]}$    	  &-0.086 &0.019 &0.033  & 6.28\\
${\rm [Zn/Fe]}$    	  &-0.035 &0.042 &0.072  & 4.59\\ 
${\rm [Y/Fe]}${\sc ii}    &-0.110 &0.038 &0.066  & 2.21\\
${\rm [Zr/Fe]}${\sc i}    &-0.177 &0.023 &0.040  & 2.60\\ 
${\rm [Zr/Fe]}${\sc ii}   &-0.177 &0.029 &0.050  & 2.60\\ 
${\rm [Ba/Fe]}${\sc ii}   & 0.047 &0.077 &0.133  & 2.18\\
${\rm [La/Fe]}${\sc ii}   & 0.120 &0.046 &0.080  & 1.10\\
${\rm [Ce/Fe]}${\sc ii}   & 0.159 &0.046 &0.079  & 1.63\\ 
${\rm [Pr/Fe]}${\sc ii}   &-0.032 &0.107 &0.185  & 0.71\\ 
${\rm [Nd/Fe]}${\sc ii}   & 0.053 &0.056 &0.097  & 1.45\\ 
${\rm [Eu/Fe]}${\sc ii}   &-0.350 &0.050 &0.087  & 0.52\\
\hline
\end{tabular}
\label{mean}
\end{table}

\begin{table}
\centering
\caption{Comparison of average abundances.}
\begin{tabular}{lrrrr}
\hline
       & here &F11 &JF &Sun lit.\\
\hline
${\rm [Fe/H] }$ &-0.06  &-0.08 &-0.04 & 7.52\\    
${\rm [Na/Fe]}$ &+0.17  & 0.05 & ...  & 6.33\\	  
${\rm [Mg/Fe]}$ &+0.07  & 0.21 & ...  & 7.58\\	  
${\rm [Si/Fe]}$ &+0.04  & 0.19 & ...  & 7.55\\	  
${\rm [Ca/Fe]}$ &+0.12  & 0.05 & ...  & 6.36\\	  
${\rm [Ti/Fe]}$ &-0.06  &-0.04 & ...  & 4.99\\	  
${\rm [Ni/Fe]}$ &-0.09  & 0.02 & ...  & 6.23\\	  
${\rm [Zr/Fe]}$ &-0.18  & 0.24 & 0.49 & 2.95\\    
${\rm [Ba/Fe]}$ &+0.05  & ...  & 0.58 & 2.31\\	  
${\rm [La/Fe]}$ &+0.12  & ...  & 0.18 & 1.21\\	  
${\rm [Eu/Fe]}$ &-0.35  & ...  &-0.02 & 0.51\\	  
\hline			
\end{tabular}
\label{conf2}
\end{table}


\subsection{Comparison with previous spectroscopy}\label{sec:comparison}

\cite{mermilliod08} used CORAVEL at the OHP (Observatoire Haute Provence) to
obtain repeated RV measures of 12 stars in NGC~2355. Four were found to be not
members and one of the eight members is a binary. The average RV of the seven
single members is $35.02\pm0.16$ (rms=0.42) km~$s^{-1}$. Table~\ref{conf} shows
the comparison of our and their RVs; they are in very good agreement. 

\begin{table}
\centering
\caption{Literature values for the three stars.}
\begin{tabular}{cccccccc}
\hline
ID   & RV  & RV  &T$_{\rm eff}$ &$\log$g &[Fe/H] \\
     &(M08)&(S00)&(S00)     &(S00)   &(S00)  \\
\hline
 201728   & 34.07 & 34.79 & 4987 & 2.72 & -0.14 \\ 
 201729   & 34.51 & 34.89 & 4961 & 2.67 & -0.15 \\ 
 201735   & 34.58 & 35.73 & 5122 & 2.73 & -0.23 \\ 
\hline			  
\end{tabular}
\begin{list}{}{}
\item[]	M08=\cite{mermilliod08}, S00=\cite{soubiran}
\end{list}
\label{conf}
\end{table}

To derive the cluster's properties, \cite{soubiran} combined spectra obtained with ELODIE on the 1.93m telescope at the OHP with the \cite{km91} photometry, 2MASS data, and proper motions. 
They obtained spectra of 24 stars, both evolved and at the MSTO; 17 turned out to be member on the basis of their RV. The eight stars near the MSTO have high rotational velocities; the average RV defined by the nine giants is
35.13 (rms=0.39) km~$s^{-1}$, also in agreement with our values. \cite{soubiran} used TGMET, a software which finds the best matching template in a library, to derive atmospheric parameters and absolute magnitude of the stars (see their Fig.~7 and Table~2). 
We observed three stars in common and Table~\ref{conf} shows that there is a satisfactory agreement between our studies. 

\cite{jacobson11} observed NGC~2355 with the multiobject spectrograph Hydra at the 3.5m WYIN telescope on Kitt Peak; the resolution was about 18000 and the wavelength coverage
about 300 \AA, centred at 6280 \AA. They obtained spectra for a dozen stars, half of which were found member (see their paper for details and for comparison with previous results, usually -but not-always- in good agreement). 
Abundance analysis was based on EWs and synthesis, using MOOG \citep{sneden73}.
Finally, \cite{jf} observed three of these stars with the echelle spectrograph on the KPNO 4m telescope.
None of our three stars is in common with these two studies.
However, the average temperatures and gravities are
4975, 2.97 and 5200, 2.95 for the four RC stars in \cite{jacobson11} and the two RC stars in \cite{jf}, respectively, compared to our average values of 5062, 2.64. 

We present in Table~\ref{conf2}  a comparison between our average abundance ratios and those in \cite{jacobson11} and \cite{jf}
for the species in common. The metallicity agrees very well, whereas we find  differences of the order of 0.1 dex for the light, $\alpha$, and iron-peak elements (given only in \citealt{jacobson11}). This is a common occurrence when comparing inhomogenoeus analyses and can be due to many factors such as different solar reference values (as in the cases of Na and Ca, where this would explain fully the offsets), different choice of lines, $\log gf$, temperatures, etc. Given the small offests, we do not explore further these cases.

Very large discrepancies are found instead for the three neutron-capture elements Zr, Ba, and Eu; conversely, there is a very good agreement for the lanthanum abundances.   This deserve investigation. Unfortunately, we have no stars in common,  thus a direct comparison is hampered.  

For Zr, we found a ratio of [Zr/Fe]=$-$0.18 to be compared to [Zr/Fe]=+0.49 by \cite{jf}, implying a difference of 0.67 dex (more than a factor of four).

We can only partially explain such a large disagreement in terms of the adopted
atomic parameters. Those authors discussed
that their oscillator strengths results in a solar Zr abundance larger
than all literature estimates by $\sim$ 0.3-0.4 dex (see that paper for details). 
However, for Zr~{\sc i} lines in common with that study the average differences in log$gf$ values are approximately 0.15 dex, so that they cannot account for the discrepancy. It is noteworthy that our Zr abundances from neutral and single-ionised lines do agree very well, suggesting that the different ionisation stage of the lines under scrutiny is not responsible for the mismatch. 
On the other hand, we note also that in \cite{jf} there is a large scatter from different
Zr~{\sc i} lines for two of the three stars, with values ranging from +0.19 to
+0.60, and from +0.59 to +0.92, respectively. This seem to suggest that the
[Zr/Fe] ratios could be quite uncertain and must be taken with caution.

The $r$-process element Eu presents a difference (in the sense ours minus
\citealt{jf}) of $-$0.33 dex, which is beyond measurement errors. While from our
data the Eu content is very homogeneous within the cluster,  \cite{jf} detected
significant internal variations, finding a difference between star \#144 and
stars \#398, 668 of a factor of two and four, respectively, in their [Eu/Fe] ratios. These differences seem due to one of the two lines they used; had they relied only on the same line we also are using (6645 \AA), results would have been much more homogenous and more similar to our values (however, this line has been measured only in two of the three stars).  Since open clusters are
not known to host any intrinsic internal variations in terms of chemical
composition, we are tempted to conclude that our analysis and the use of the best line is more robust. 

Finally, given its peculiar pattern observed in open cluster stars, Ba deserves a more detailed discussion. 
First identified by \cite{dorazi09}, and subsequently confirmed by several different studies  (e.g., \citealt{jf}, see discussion in Section 6), there is a trend of decreasing Ba abundances as a function of the open cluster's age, with younger clusters exhibiting [Ba/Fe] ratios up to $\sim$ 0.6 dex (such as e.g., the case of IC~2602 and IC~2391 by \citealt{dorazi09}), whereas solar ratios are measured for clusters a few Gyr old. Considering the age of NGC~2355, and values reported in the literature for almost coeval clusters, we should expect enhancements in the Ba content at levels of approximately 0.25$-$0.40 dex, which is not evident from the present analysis. 
In fact, our average mean abundance results in a solar [Ba/Fe] ratio of 0.05$\pm$0.08 dex (0.53 dex lower than Jacobson's study). The different techniques and approaches could in principle explain this large discrepancy; for instance, equivalent width analysis has been carried out by \citealt{dorazi09} and \citealt{jf}, whereas here we performed spectral syntheses. Critical in this respect is also the microturbulence value, because Ba~{\sc ii} lines sit on the saturated part of the curve of growth and they are quite insensitive to the abundance. Given the quite limited sample, we cannot certainly state that this cluster presents an anomalously low Ba content 
(considering its age) or, conversely, measurement errors cause this trend. Homogeneous analysis of all the data, or of a similarly large sample of clusters,  should  be attempted, to clarify this issue.

\section{Synthetic CMDs} \label{sec:synth}
To complete our analysis of NGC~2355, the age, distance modulus,  reddening (total and differential), and
binary fraction of the cluster are estimated using the synthetic CMD technique \citep[see][]{tosi91} as done in all the papers of the
BOCCE project (see e.g., \citealt{cignoni11,donati12,lbt2} and references therein). Homogeneous sets of three types of stellar evolution models\footnote{The Padova \citep{bbc93}, FRANEC
\citep{franec}, and FST \citep{fst} with $\eta=0.2$. We keep using these relatively old models for sake of homogeneity within the BOCCE sample of clusters.} are used to build a library of synthetic CMDs. Cluster parameters are determined by means of the comparison of the synthetic CMDs with the observed ones. The best-fitting solution is chosen as the one that can best reproduce age-sensitive indicators: the MSTP, the RC luminosity, the MS inclination
and colour, and the RC colour. In order to make a meaningful comparison, the synthetic CMDs are made taking into account the photometric error, the completeness level of the photometry and the stellar density contrast of the OC population with respect to the population of an external field. The synthetic CMDs are combined with stars picked from an equal area of the external field to take the contamination into
account. 

Multi-colour photometry has generally proven to be fundamental to obtain the best parameters estimation. Hence, the best-fitting solutions are the ones that can reproduce at the same time all the observed CMDs (in our case,  $V,B-V$ and $V,V-I$) for appropriate distance modulus, reddening, metallicity, and age.  For NGC~2355 we fixed the metallicity  of the models to the spectroscopic one (i.e. we adopted the evolutionary tracks with solar value: $Z=0.02$), which helped us restricting the possible range of parameters.  With this metallicity we obtained a good fit of both $V,B-V$ and $V,V-I$ observational CMDs after adopting the standard extinction law ($E(V-I)=1.25\times E(B-V)$, $R_V=3.1$, see \citealt{dean_78}), circumstance that further confirms the spectroscopic metallicity and indirectly supports the accuracy of our calibration. 

\subsection{Cluster parameters}
The best solution for each set of tracks is the one whose synthetic CMD fits most of the visible MS shape, the RC and MSTP levels, the binary sequence, and, if present, the broadening of the MS due to differential reddening. In general we found RC colour sligthly redder in the synthetic CMDs than in the observed one, especially in the $V,B-V$ case. However,  this evolutionary stage strongly depends on fundamental parameters (e.g. metallicity, age, helium content), and on physical inputs (e.g. efficiency of core overshooting, mass loss) to which colour is very sensitive. We also found an overall good agreement for the MS in the $V,V-I$ CMD and a worse fit in the case of the $V,B-V$ one, where the correct inclination of the sequence is not well reproduced. This is a rather common occurrence and probably the main driver of this mismatch is
the uncertainty in the adopted temperature-colour transformations and  in the model atmospheres \citep[as discussed in ][]{boc_06}. However, we can not exclude a priori that systematic errors might be hidden in the calibration of the photometry because we could use only tertiary standards (see Sec.~\ref{sec:calibration}). Concerning the fraction of binary systems and differential reddening we fine-tuned our synthetic CMDs for different amount of these two quantities (as done in other papers of this series, see for example \citealt{donati12,lbt2}). 
\begin{figure}
\includegraphics[scale=0.45]{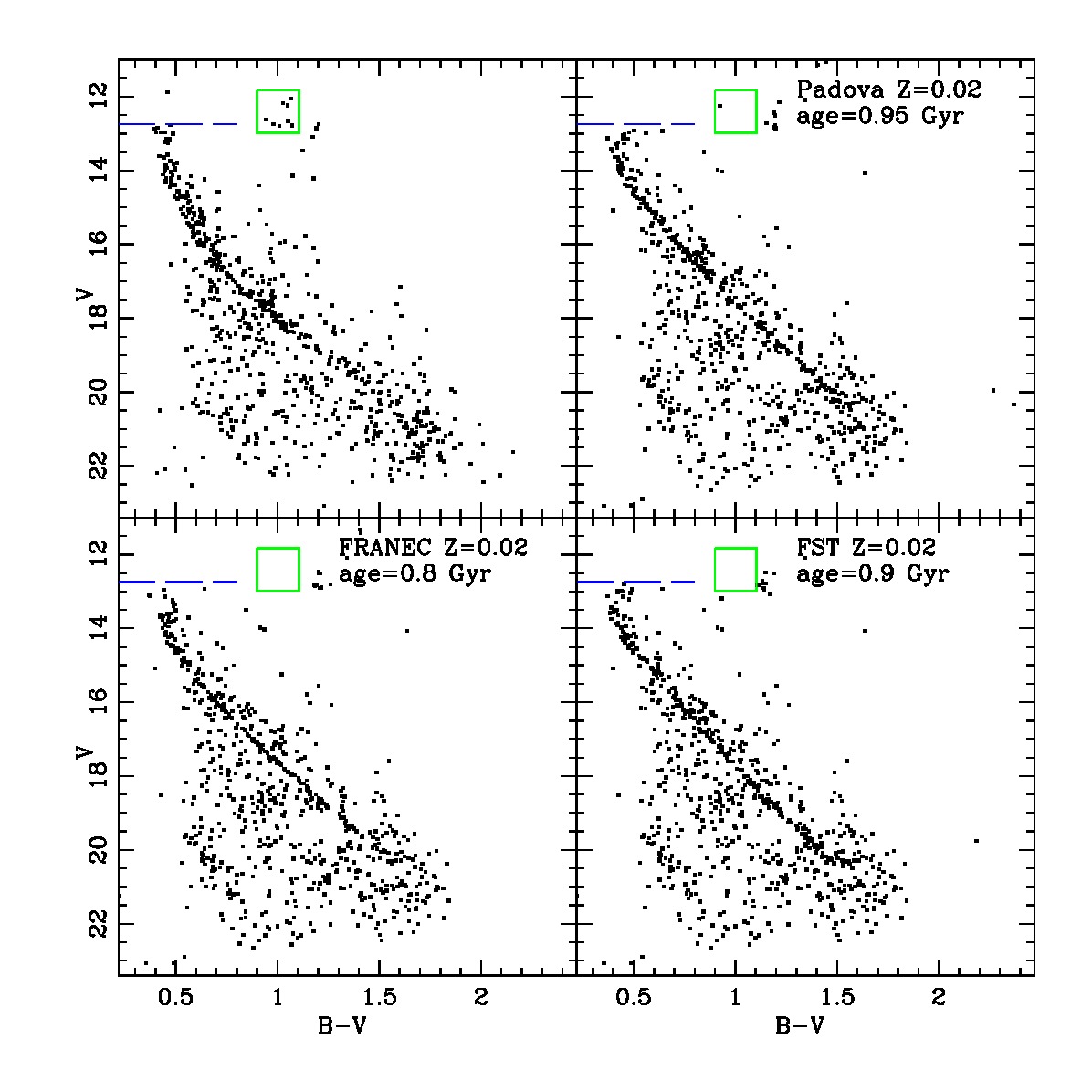}
\includegraphics[scale=0.45]{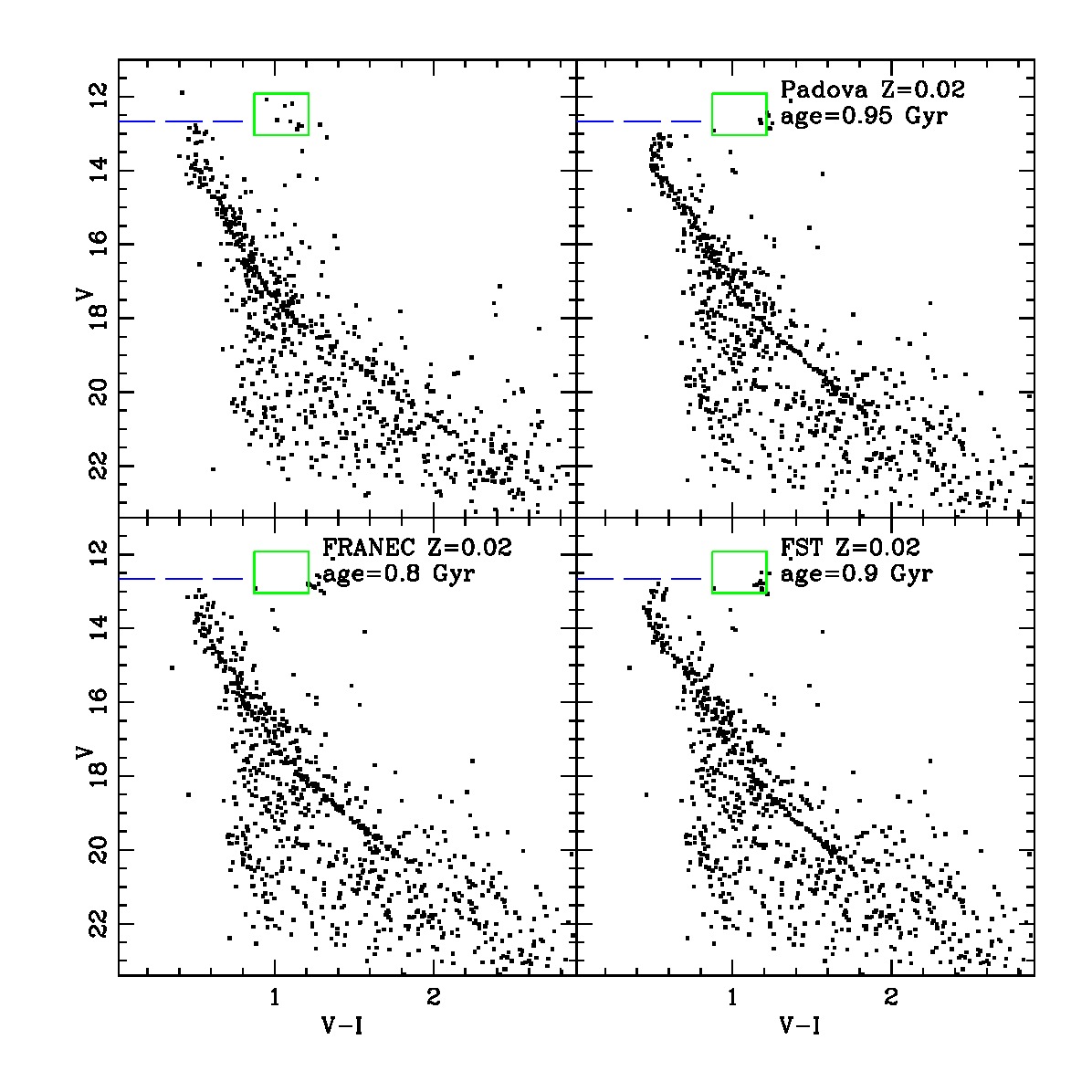}
\caption{In both boxes the upper left CMD is the observational one. The other CMDs are the synthetic ones for the different evolutionary models labelled in each panel and described in this section.}
\label{figsynths}
\end{figure}
For binaries, we build synthetic CMDs where the desired percentage of objects is not a single star but objects that have the photometric properties of two randomly picked synthetic stars as they were photometrically unresolved. In the case of differential reddening, the adopted quantity is considered as an upper limit and added as a random positive constant to the mean Galactic reddening. We found that the best estimation for the binary fraction is 35\% and for the differential reddening is 0.08 mag, and kept them fixed them for all the following analysis.  

Figure~\ref{figsynths} shows the comparison between the observational CMD inside 5$\arcmin$ from the cluster centre (a good compromise between having enough cluster stars in all evolutionary phases and low field stars contamination) and the best fits obtained with the three sets of tracks. We recall that solar metallicity is assumed in all tests. To better compare data and models we draw the age sensitive indicators (see Sec.~\ref{sec:CMD}) both in the observed and in the synthetic CMDs.

With the Padova models we found an age of 0.95 Gyr, $E(B-V)=0.14$ mag, and distance modulus $(m-M)_0=10.92$ mag. This model can adequately reproduce the luminosity levels of the MSTP and RC. For the latter we obtained a good match of the colour on the $V,V-I$ plane while in $V,B-V$ the synthetic clump is too red (about 0.1 mag redder). Especially in the $V,V-I$, the model well fit the observed MS down to magnitude $V\sim$ 20 mag.

For the FST models we found the best fit for an age of 0.9 Gyr, $E(B-V)=0.15$ mag, and distance modulus $(m-M)_0=11.04$ mag. The colour and magnitude of the MSTP are well reproduced and we found a very good fit to the MS especially in $V,V-I$ down to magnitude $V\sim19$ mag. The luminosity and colour of the RC are perfectly reproduced in the $V,V-I$ plane while the synthetic RC is slightly redder than observed in the $V,B-V$ plane.

In the case of the FRANEC models we obtained the best fit for an age of 0.8 Gyr, $E(B-V)=0.19$ mag, and distance modulus $(m-M)_0=10.91$ mag. For this set of models we obtained the best fit of the MS down to magnitude $V\sim22$ mag, in particular in the $V,V-I$ plane, while the RC colour is too red in both $V,B-V$ and $V,V-I$ synthetic CMDs. As expected, the ages derived from the FRANEC models are lower than those from the other models. This is because the FRANEC tracks do not include overshooting from convective cores, while the other two sets do.

Table~\ref{tab:summary} shows the cluster parameters we derived, together with
the implied Galactocentric distance, height above the Galactic plane, and mass at the MSTO.
The typical errors on the cluster parameters are of the order of 0.05 Gyr in age, 0.05 mag in reddening and 0.1 mag in distance modulus as found in other papers of this series. 
For comparison, literature values range between 0.7 and 1 Gyr for the age, 0.1 and 0.3 mag for the reddening, 10.8 and 12 mag for the distance modulus (see Tab.~\ref{t:lit}). The age estimates are in good agreement with our analysis while the broad literature range of the other two parameters is likely due to the intrinsic difference of the analysis method and to the different metallicity estimates adopted (usually lower than the solar value we used). In fact, metallicity has a larger impact on reddening and distance modulus than on age, which is less effected.  

\begin{table*}
  \centering
  \caption{Cluster parameters derived using different models. Recall that the spectroscopic metallicity we found is slightly lower than solar.}
  \begin{tabular}{|l|c|c|c|c|c|c|c|c|}
    \hline
    \hline
 Model & age & $Z^a$ & $(m-M)_0$ & $E(B-V)$ & $d_{\odot}$ & $R_{GC}^b$ &Z & $M_{TO}$\\
       & (Gyr) &     & (mag)     & (mag)    & (kpc)       & (kpc)      & (pc) & ($M_{\odot}$)\\
    \hline
 Padova & 0.95 & 0.02 & 10.916 & 0.14 & 1.52 & 9.39 & 312.1 & 2.1 \\
 FST    & 0.9  & 0.02 & 11.035 & 0.15 & 1.61 & 9.47 & 329.7 & 2.2 \\
 FRANEC & 0.8  & 0.02 & 10.911 & 0.19 & 1.52 & 9.39 & 311.4 & 2.2 \\ 
    \hline
  \multicolumn{9}{l}{$^a$Metal content of the evolutionary tracks.}\\
  \multicolumn{9}{l}{$^bR_{\odot}=8$ kpc is used to compute $R_{GC}$.}\\
  \end{tabular}
  \label{tab:summary}
\end{table*}

\begin{figure*}
\centering
\includegraphics[scale=0.8]{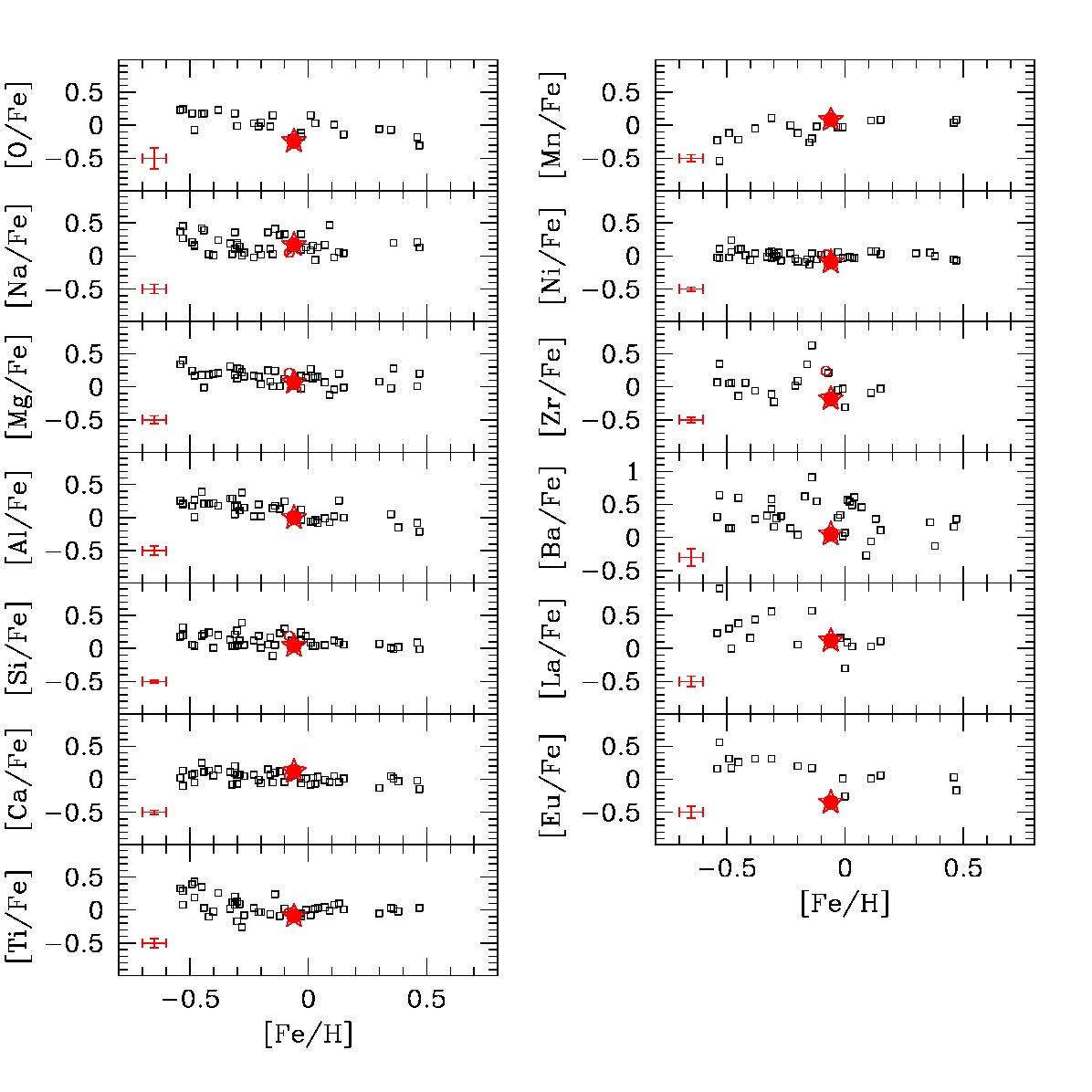}
\caption{Comparison of results for NGC~2355 (indicated by a filled red star) and literature values for open clusters (with multiple results for the same cluster all shown, see text) for the elements in common. The errorbars in each panel are the rms of our values. }
\label{yong}
\end{figure*}

\begin{figure*}
\centering
\includegraphics[scale=0.8]{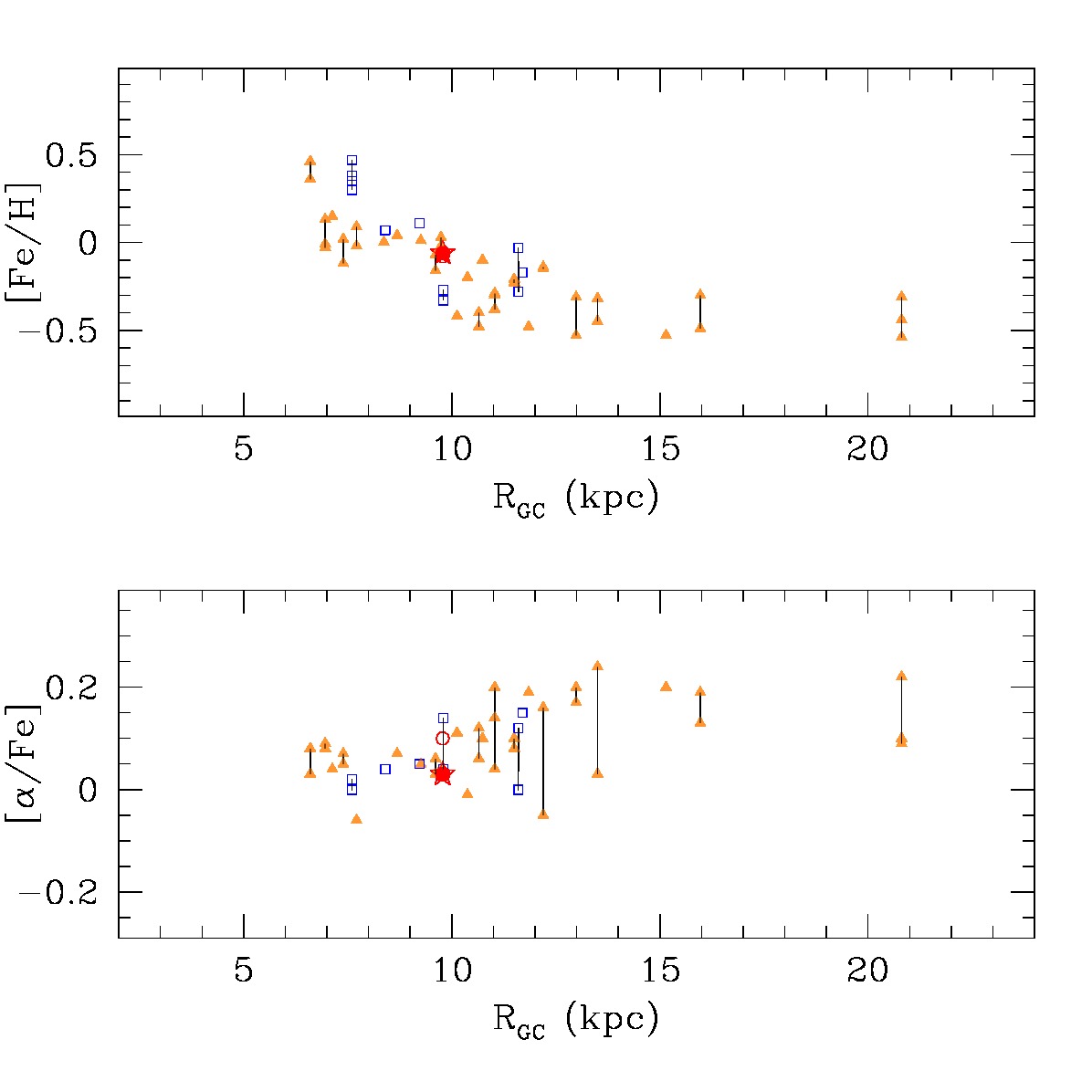}
\caption{Metallicity and $\alpha$-elements distributions with Galactocentric distance. Our value for NGC~2355 is indicated by a red star. Orange filled points are clusters with BOCCE age and
$R_{GC}$, open blue squares clusters with literature age and $R_{GC}$. Lines connect different metallicity and [$\alpha$/Fe] values for the same cluster. References for the spectroscopic and photometric
papers are in Table~13.}
\label{grad}
\end{figure*}

\begin{figure*}
\includegraphics[scale=0.85]{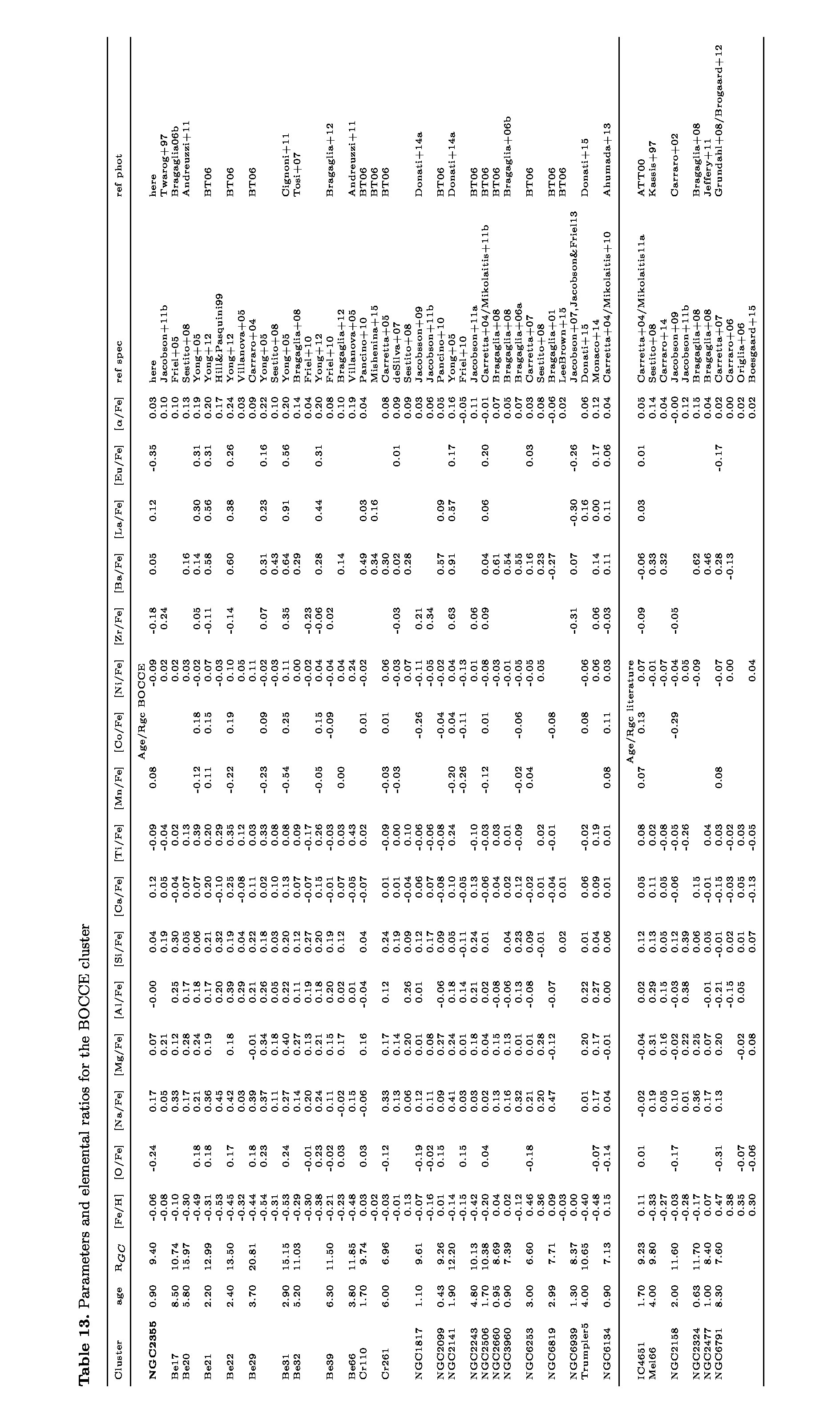}
\label{tablegrad}
\end{figure*}

\section{Discussion}

With the analysis of NGC~2355, both with photometry, to obtain age, distance, and reddening, and with spectroscopy for detailed chemistry, we add another cluster to the homogenous sample of BOCCE clusters.

To put NGC~2355 in the context of other OCs studied, we plot in
Figure~\ref{yong} our results on a compilation of literature data. We took
Table~13 of \cite{Yong+12} as a starting point, but added other clusters analyzed in later papers. Then, to have a more homogenous set, we selected only clusters in the BOCCE sample, either because we derived their distance and age (and often, but not necessarily, the chemical abundances) or their composition (but not age and distance).  We ended up with 30 selected clusters, for six of which we adopted literature distance and age while all the others have our BOCCE values; 12 of the 30 clusters have only one spectroscopic analysis, 14 have two, and four have three or more.
NGC~2355 falls within the distribution of the other clusters abundances. The information on this sample of OCs (clusters, distance, age, metallicity, and selected elemental ratios, and references for these values) is given in Table 13. No homogenisation has been done, we report results from the original papers.

Classically, OCs have been used to describe the disc metallicity distribution
\citep[e.g.,][to cite only a few papers]{fri_95,magrini09,lepine,andreuzzi,Yong+12,heiter}. The need of a homogeneous analysis for a large sample of clusters is one of the drivers of the BOCCE project; however, we have not yet reached that goal. Large surveys
(e.g., Gaia-ESO, APOGEE, GALAH, see \citealt{gilmore12,apogee,galah}, respectively) targeting also clusters will be fundamental in this respect, but
they have not reached yet the ``critical mass". In the meantime, Fig.~\ref{grad} shows the possibilities and the limits of what can be done with the present samples (see also the discussion in \citealt{heiter}). In the upper panel we show the metallicity distribution, which displays a negative radial gradient in most of the disc, but a flatter slope in the outer part ($R_{GC} \ga 12$ kpc)\footnote{All $R_{GC}$ values have been computed using uniformly $R_{GC,\odot}=8$ kpc, also for the clusters  for which we took distances from literature. }. Lines connect the different metallicities for the same cluster and we see that in some cases the differences exceed 0.1 dex, a reasonable average errorbar for the measures. The radial metallicity distribution of OCs can be interpreted in the light of chemical evolution models (see e.g. \citealt{chiappini01,magrini09,romano10}), in which the rapid star formation rate in the inner disc together with the assumption of a different rate of infalling material from  the intergalactic medium onto the the inner and outer disc, predict a negative metallicity gradient in the inner disc and a flattening in the outer one. However, the
issue of the time evolution of the radial abundance gradients is far from being settled 
unequivocally both from the theoretical (different model prescriptions lead to different scenarios: either a flattening or a steepening of the inner gradient with time) and observational points of view (some authors suggest that the radial metallicity distribution is not a negative gradient but rather a step distribution, see e.g. \citealt{twarog97,lepine}). 
The interpretation becomes even more intriguing when dynamics is considered in galactic chemical evolution models. For example, flat abundance gradients could also be produced when radial mixing is taken into account \citep[e.g.,][]{roskar08,sanchez09,minchev11,minchev12}. \cite{fuji12}, using N-body simulations, found a time-scale of 100Myr for the radial migration of OCs. They demonstrated that in this time a cluster could move 1.5 kpc away from its birth location, a result stressing how important the dynamics can be in interpreting the chemical evolution history of the Galactic disc. 
However, \cite{haywood13}, by analyzing FGK stars near the Sun, argue for the lack of a detectable influence of radial migration, at least in the solar neighborhood.
Determinations of abundance gradients in the disc at different ages are thus crucial to disentangle among different evolutive scenarios. The Gaia-ESO Survey\footnote{http://www.gaia-eso.eu/}, with its large spectroscopic legacy (high and intermediate resolution spectra of about $10^5$ stars in the MW and stars in about 70 OCs) will have an important impact on this side. Furthermore, within the Survey there are ongoing projects focussed on the accurate analysis of clusters orbits (e.g. Jacobson et al. in prep.), following the same approach described in \cite{jilkova12} in the case of NGC~6791, with the aim to build a comprehensive study of OCs which takes into account homogeneous chemical abundances and kinematics.
Gaia\footnote{http://sci.esa.int/gaia/} will further improve these studies with precise proper motions, helping in determining the impact of dynamics on the MW evolution.

In the lower panel of Fig.~\ref{grad} we show, for the same sample of OCs, the run of $\alpha$-elements with $R_{GC}$. The distribution mirrors that of \cite{Yong+12} (see their Fig.~21), and seems to indicate that inner and outer disc OCs
have a different level of $\alpha$ enrichment, with a difference of 0.10-0.15 dex, again considering clusters within about 12 kpc from the Galactic centre and farther than that. However, when discussing both the distributions shown in Fig.~\ref{grad}, we have to keep in mind that the abundance analysis is not fully homogenous. As an example of the importance of a homogenous analysis, we recall the recent study of \cite{magrini15} where four inner disc OCs were studied, also in comparison with field stars analysed in the same way, using the same line lists, reference values, etc. In this way it was possible to highlight subtle differences in their chemical composition, being sure that they were not due to systematic errors, and question their origin within the Galactic disc. Similar, more complete studies will be performed on the complete OCs database of the Gaia-ESO Survey and of other similar surveys

\begin{figure}
\centering
\includegraphics[scale=0.8]{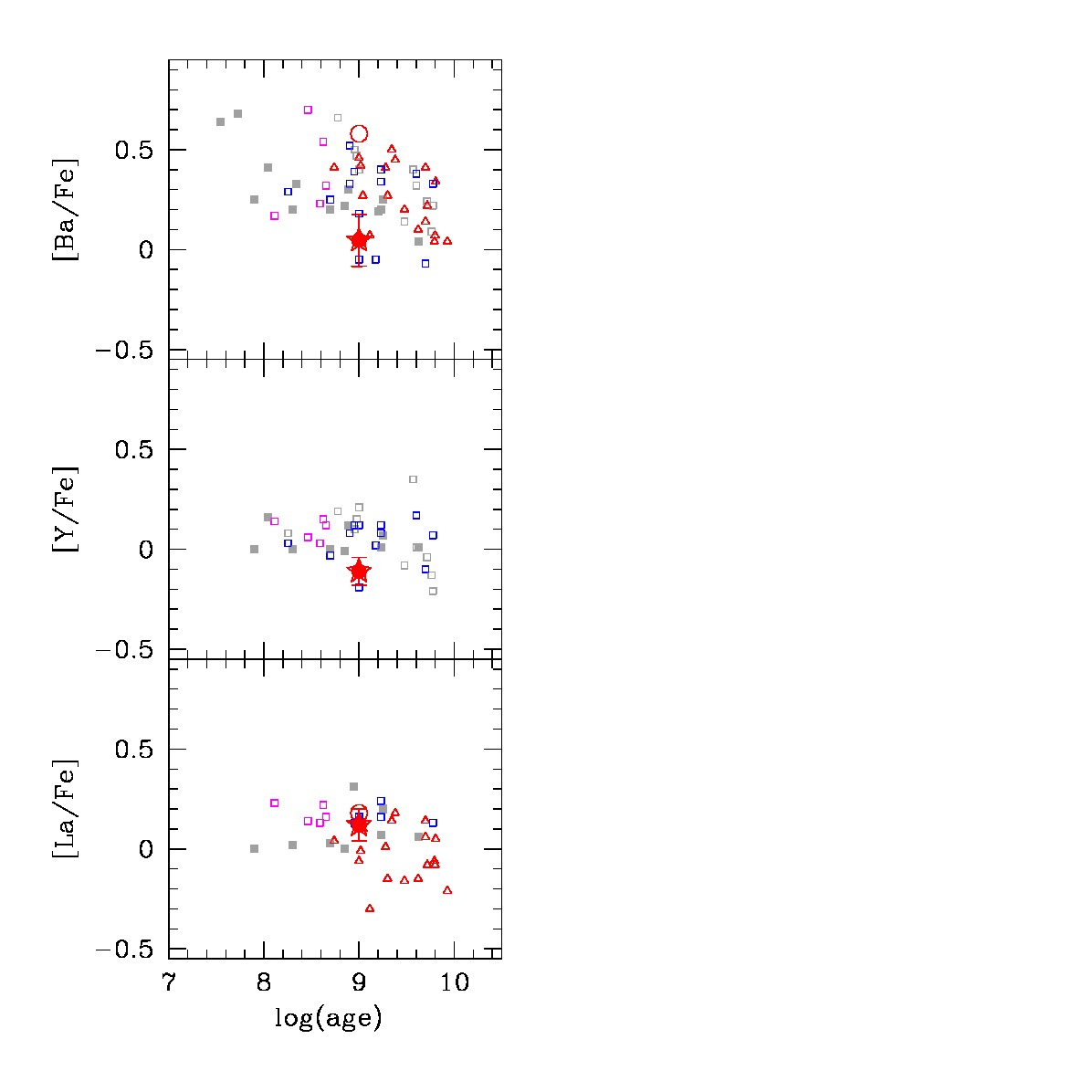}
\caption{Run of [Ba/Fe], [La/Fe], and [Y/Fe] with age obtained by three
homogeneous analyses (D'Orazi et al.
2009, 2012, Maiorca et al. 2011: open grey squares for giants, filled grey squares for dwarfs; Jacobson \& Friel 2011: open  red triangles; Mishenina et al. 2013, 2015: open blue squares; in both cases only giants). For NGC~2355
the value by Jacobson \& Friel is indicated by a red empty circles, our by a red star (with errorbars indicating the rms of the measure, see Table~\ref{mean}).}
\label{neutron}
\end{figure}

Finally, Figure~\ref{neutron} shows the run of the abundance ratios [Ba/Fe], [Y/Fe],
and [La/Fe] with cluster age; in the figure we use our values for NGC~2355 and three sets of data, taken from papers devoted to the study of neutron capture elements in large samples of clusters; a) \cite{dorazi09,maiorca11,dorazi12}; b) \cite{jf}; c) \cite{mish13,mish15}. We took the original values from the papers, without any attempt at homogeneization; we used the BOCCE value for age, if available (but taking the original value would not change the picture). 
As mentioned in Section~\ref{sec:comparison}, it is now well established that OCs do exhibit an anticorrelation between the [Ba/Fe] ratios and the clusters' age. In order to provide a theoretical explanation to this abundance trend, \cite{dorazi09} (and subsequently \citealt{maiorca11,maiorca12}) suggested that Ba production in low-mass AGB stars (M $<$ 1.5 M$_{\odot}$) should be more efficient than that predicted by standard chemical evolution models, being the $^{13}$C pocket, that is responsible for providing neutrons, larger than previously thought. If this were the case, then all the $s$-process elements, and in particular La and Ce (belonging to the same second s-process peak) should follow the Ba pattern, though at different extent. However, different studies have provided contradictory results: whereas \cite{maiorca11} concluded that all the other s-process elements reflect this increasing trend with decreasing age, \cite{dorazi12}, \cite{Yong+12}, 
\cite{mish13}, and very recently \cite{reddy15} did not confirm such finding. This is also evident in Figure~\ref{neutron}, where Y and La, at variance with Ba, show a quite flat behaviour with the OC age. To draw final conclusions on this issue, larger and very homogeneous samples are needed, especially for clusters younger than roughly 500 Myr; at the moment several possibilities remain open (we refer the reader to \citealt{dorazi12} for further discussion on this topic, including non-LTE and activity effects impacting on Ba abundances). It is noteworthy that, since s-process models cannot account for Ba production without bearing enhancements in the other s-process species, \cite{mish15} suggested a different synthesis channel to explain this peculiar trend, the so-called intermediate ($i$) n-capture mechanism where neutron densities are intermediate between the low densities of the $s$-process and the very high values of the r-process (see \citealt{cowan}).

\section{Summary}
We were able to perform a complete analysis of the parameters of NGC~2355, combining precise photometry from LBC@LBT and high-resolution spectroscopy with FIES@NOT.

We obtained observational CMDs two magnitude deeper than literature ones and on a larger FoV. By means of the well tested analysis of cluster parameters with the synthetic CMD technique, we derived age, distance, reddening, differential reddening, and binary fraction with three different set of tracks (Padova, FST, and FRANEC). We found that NGC~2355 is located at about 1.6 kpc from the Sun. Its Galactic position, towards the anti-centre, is at $R_{GC}\sim9.4$ kpc and 300 pc above the plane (assuming $R_\odot=8$ kpc). The age is between 0.8 and 0.95 Gyr, depending on the adopted stellar model. The mean Galactic reddening is $0.14<E(B-V)<0.19$ mag, while the estimated differential reddening is about 0.08 mag and the estimated fraction of binaries is about 35\%. 

The analysis of the high-resolution FIES spectra of three giant stars (belonging to the RC) suggests an almost solar metallicity, with an average [Fe/H]$=-0.06\pm0.04$ dex. We derived abundances for O, the light elements Na and Al, the $\alpha$-process elements Mg, Si, Ca, Ti (from Ti I and II lines), the Fe-group elements Sc, Cr (from neutral and singly-ionised features), Mn, Ni, Cu, Zn, and for the neutron-capture elements Y, Zr, Ba, La, Ce, Pr, Nd, Eu.
When comparing with other OCs at the same metallicity and Galactocentric distance, NGC~2355 falls within the distribution of the other cluster abundances. However, there are some  exceptions, especially for neutron capture elements. We discussed possible explanations for these differences, in particular for Ba, which resulted low for the cluster age, compared to expectations based on previous studies. In this case, further investigations are needed to reach a firm conclusion which is, at the moment, constrained only by a limited number of stars.

This analysis represents another step towards the homogeneous tracing of the chemical properties and evolution of the Galactic disc set by the BOCCE project. We are aiming at a database of about 50 OCs to reach statistically significant and accurate results in order to shed a more robust light on the apparent uncertainties that still drive the interpretation of the Galctic disc properties. The achievements of this project, combined with the unprecedented spectroscopic legacy of surveys such as Gaia-ESO or APOGEE and the forthcoming results of the astrometric mission Gaia, will have an important impact on our understanding of the MW.

\section*{Acknowledgements}
We thank Paolo Montegriffo, whose software for catalogue matching we
consistently use for our work. We thank the personnel at LBT and NOT for their help and for taking part of the data. For this paper we used  the VizieR catalogue
access tool (CDS, Strasbourg, France), WEBDA (originally created by J.-C. Mermilliod and now maintained by
E. Paunzen), and NASA's Astrophysics Data
System. This research has been partially funded by MIUR and INAF (grants ``The Chemical
and Dynamical Evolution of the Milky Way and Local Group Galaxies'' ,
prot. 2010LY5N2T; grant ``Premiale VLT 2012").
Funding for the SDSS and SDSS-II has been provided by the Alfred P. Sloan Foundation, the Participating Institutions, the National Science Foundation, the US Department of Energy, the NASA, the Japanese Monbukagakusho, the Max Planck Society, and the Higher Education Funding Council for England. The SDSS website is http://www.sdss.org/. The SDSS is managed by the Astrophysical Research Consortium for the Participating Institutions. The Participating Institutions are the American Museum of Natural History, Astrophysical Institute Potsdam, University of Basel, University of Cambridge, Case Western Reserve University, University of Chicago, Drexel University, Fermilab, the Institute for Advanced Study, the Japan Participation Group, Johns Hopkins University, the Joint Institute for Nuclear Astrophysics, the Kavli Institute for Particle Astrophysics and Cosmology, the Korean Scientist Group, the Chinese Academy of Sciences (LAMOST), Los Alamos National Laboratory, the Max-Planck-Institute for Astronomy (MPIA), the Max-Planck-Institute for Astrophysics (MPA), New Mexico State University, Ohio State University, University of Pittsburgh, University of Portsmouth, Princeton University, the United States Naval Observatory, and the University of Washington.

\bsp

\label{lastpage}

\end{document}